\documentclass{nature}

\usepackage{graphicx}
\makeatletter
\let\saved@includegraphics\includegraphics
\AtBeginDocument{\let\includegraphics\saved@includegraphics}
\renewenvironment*{figure}{\@float{figure}}{\end@float}
\makeatother

\usepackage{amsmath,amstext}

\usepackage{xr}
\usepackage{hyperref}

\newcommand\msunyr{$M_\odot$ yr$^{-1}$}

\newcommand{\beginmethods}{%
  \setcounter{table}{0}
  \renewcommand{\thetable}{\arabic{table}}%
  \renewcommand{\tablename}{Supplementary Table}
  \setcounter{figure}{0}
  \renewcommand{\thefigure}{\arabic{figure}}%
  \renewcommand{\figurename}{Supplementary Figure}
  \setcounter{section}{0}
  \renewcommand{\thesection}{\arabic{section}}%
}

\title{The delay of shock breakout  due to circumstellar material seen in most
  Type II Supernovae\footnote{Published version in \url{https://www.nature.com/articles/s41550-018-0563-4}}
}

\begin{document}

\maketitle 

\vspace{-1cm}

\paragraph*{Authors:} F.~F\"orster$^{1, 2, 3\ast}$,
      T.~J.~Moriya$^{4}$,
      J.C.~Maureira$^{1}$,
      J.P.~Anderson$^{5}$,
      S.~Blinnikov$^{6, 7, 8}$,
      F.~Bufano$^{9}$,
      G.~Cabrera--Vives$^{10, 2}$,
      A.~Clocchiatti$^{11, 2}$, 
      Th.~de Jaeger$^{12}$,
      P.A. Est\'evez$^{2, 13}$,  
      L.~Galbany$^{14}$,
      S. Gonz\'alez--Gait\'an$^{1, 15}$,
      G. Gr\"afener$^{16}$,
      M.~Hamuy$^{2, 3}$, 
      E.~Hsiao$^{17}$,
      P.~Huentelemu$^{13}$,
      P.~Huijse$^{2, 18}$, 
      H.~Kuncarayakti$^{19, 20}$,
      J.~Mart\'inez-Palomera$^{3, 2, 1}$,
      G.~Medina$^{3}$,
      F.~Olivares E.$^{2, 3}$, 
      G.~Pignata$^{21, 2}$,
      A.~Razza$^{3, 5}$,
      I.~Reyes$^{13, 2}$,
      J.~San Mart\'in$^{1}$,
      R.C.~Smith$^{22}$,
      E.~Vera$^{1}$,
      A.K. Vivas$^{22}$,
      A.~de Ugarte Postigo$^{23, 24}$,
      S.-C.~Yoon$^{25, 26}$,
      C.~Ashall$^{27}$,
      M.~Fraser$^{28}$,
      A.~Gal--Yam$^{29}$,
      E.~Kankare$^{30}$,
      L.~Le~Guillou$^{31}$,
      P.A.~Mazzali$^{27, 32}$,
      N.A.~Walton$^{33}$,
      D.R.~Young$^{30}$\\
\small{$^{1}$Center for Mathematical Modeling, University of Chile;}
\small{$^{2}$Millennium Institute of Astrophysics, Chile;}
\small{$^{3}$Department of Astronomy, University of Chile;}
\small{$^{4}$Division of Theoretical Astronomy, National Astronomical Observatory of Japan, National Institutes of Natural Sciences, 2-21-1 Osawa, Mitaka, Tokyo 181-8588, Japan}
\small{$^{5}$European Southern Observatory, Alonso de C\'ordova 3107, Casilla 19, Santiago, Chile;}
\small{$^{6}$Kavli Institute for the Physics and Mathematics of the Universe (WPI), University of Tokyo, Japan;}
\small{$^{7}$Institute for Theoretical and Experimental Physics (ITEP), Moscow, Russia;}
\small{$^{8}$All-Russia Research Institute of Automatics (VNIIA), Moscow, Russia}
\small{$^{9}$INAF - Astrophysical Observatory of Catania, Italy;}
\small{$^{10}$Department of Computer Science, University of Concepci\'on, Chile;}
\small{$^{11}$Department of Physics and Astronomy, Universidad Cat\'olica de Chile, Santiago, Chile}
\small{$^{12}$Department of Astronomy, University of California Berkeley, USA;}
\small{$^{13}$Department of Electrical Engineering, University of Chile;}
\small{$^{14}$Department of Physics and Astronomy, University of Pittsburgh, USA;}
\small{$^{15}$CENTRA, Instituto T\'ecnico Superior, Universidade de Lisboa, Portugal;}
\small{$^{16}$Argelander-Institut fur Astronomie, Auf dem H\"ugel 71, D-53121 Bonn, Germany;}
\small{$^{17}$Department of Physics, Florida State University, USA;}
\small{$^{18}$Informatics Institute, Universidad Austral de Chile;}
\small{$^{19}$Finnish Centre for Astronomy with ESO (FINCA), University of Turku, V\"{a}is\"{a}l\"{a}ntie 20, 21500 Piikki\"{o}, Finland;}
\small{$^{20}$Tuorla Observatory, Department of Physics and Astronomy, University of Turku, V\"{a}is\"{a}l\"{a}ntie 20, 21500 Piikki\"{o}, Finland}
\small{$^{21}$Departamento de Ciencias Fisicas, Universidad Andres Bello, Avda. República 252, Santiago, Chile}
\small{$^{22}$Cerro Tololo Interamerican Observatory, La Serena, Chile;}
\small{$^{23}$Instituto de Astrof\'isica de Andaluc\'ia (IAA-CSIC), Glorieta de la Astronom\'ia, s/n 18008 Granada, Spain}
\small{$^{24}$Dark Cosmology Centre, Niels Borh Institute, University of Copenhagen, Juliane Maries Vej 30, 2100 Copenhagen \O, Denmark}
\small{$^{25}$Department of Physics and Astronomy, Seoul National University, Gwanak-ro1, Gwanakgu, Seoul 08826, South Korea}
\small{$^{26}$Monash Centre for Astrophysics, School of Physics and Astronomy, Monash University, VIC 3800, Australia}
\small{$^{27}$Astrophysics Research Institute, Liverpool John Moores University, IC2, Liverpool Science Park, 146 Brownlow Hill,  Liverpool L3 5RF, UK}
\small{$^{28}$School of Physics, O’Brien Centre for Science North, University College Dublin, Belfield, Dublin 4, Ireland.}
\small{$^{29}$Department of Particle Physics and Astrophysics, Weizmann Institute of Science, Rehovot 76100, Israel.}
\small{$^{30}$Astrophysics Research Centre, School of Mathematics and Physics, Queens University Belfast, Belfast BT7 1NN, UK}
\small{$^{31}$Sorbonne Université, Univ Paris 06 (UPMC), Université Paris Diderot, CNRS, IN2P3, UMR 7585, LPNHE, Laboratoire de Physique Nucléaire et de
Hautes Energies, Paris, France}
\small{$^{32}$Max-Planck-Institut fur Astrophysik, Karl-Schwarzschild-Str. 1, 85748 Garching bei Munchen, Germany}
\small{$^{33}$Institute of Astronomy, University of Cambridge, Madingley Road, Cambridge CB3 0HA, UK}

\vspace{1cm}

{\bf Type II supernovae (SNe) originate from the explosion of
  hydrogen--rich supergiant massive stars. Their first electromagnetic
  signature is the shock breakout, a short--lived phenomenon which can
  last from hours to days depending on the density at shock
  emergence. We present 26 rising optical light curves of SN II
  candidates discovered shortly after explosion by the High cadence
  Transient Survey (HiTS) and derive physical parameters based on
  hydrodynamical models using a Bayesian approach. We observe a steep
  rise of a few days in 24 out of 26 SN II candidates, indicating the
  systematic detection of shock breakouts in a dense circumstellar
  matter consistent with a mass loss rate $\dot{M} > 10^{-4}$
  \msunyr\ or a dense atmosphere. This implies that the characteristic
  hour timescale signature of stellar envelope SBOs may be rare in
  nature and could be delayed into longer--lived circumstellar
  material shock breakouts in most Type II SNe.}
 
\vspace{1cm}

With a new generation of large etendue facilities such as iPTF
\cite{2009PASP..121.1395L}, SkyMapper \cite{2007PASA...24....1K},
Pan--STARRS \cite{2010SPIE.7733E..0EK}, KMTNET
\cite{2011SPIE.8151E..1BK}, ATLAS \cite{2018PASP..130f4505T}, DECam
\cite{2015AJ....150..150F}, Hyper Suprime--Cam \cite{2010AIPC.1279..120T},
ZTF \cite{2017NatAs...1E..71B} or LSST \cite{2009arXiv0912.0201L} the study of rare and short--lived
phenomena in large volumes of the Universe is becoming possible. This
allows not only finding new classes of events, but also systematically
studying short--lived phases of evolution in known astrophysical
phenomena, such as SN explosions. In this work we present 26 rising
optical light curves from Type IIP/L SN candidates discovered shortly
after explosion by the High cadence Transient Survey (HiTS)
\cite[hereafter F16]{2016ApJ...832..155F}, and a systematic study of
their physical properties.

Type IIP/L SNe, hereafter SNe II, are thought to originate after the
core collapse of massive red supergiant stars (RSGs) with main
sequence masses between 8 $M_\odot$ and 16.5 $M_\odot$
\cite{2009MNRAS.395.1409S}, although see
\cite{2018MNRAS.474.2116D}. In the currently favored scenario, the
neutrino--driven mechanism (see \cite{2016ARNPS..66..341J} and
references therein), a small fraction of the gravitational energy lost
during collapse is transferred to the outer layers of the star via
neutrino heating. This triggers a shock wave which, depending on the
mass of the progenitor and the total input energy, can sometimes
unbind the envelope of the star in a SN explosion. This shock wave is
radiation dominated, Compton scattering mediated and propagates
supersonically, typically at tens of thousands of km/s
\cite{1978ApJ...225L.133F}. The shock precursor has a characteristic
optical depth, estimated equating the timescales for photons to
diffuse out of the shock front (diffusion timescale) and the timescale
for the shock to move into a new region of the star (advection
timescale), of $\tau \approx c/v_{\rm shock}$, where $c$ is the speed
of light and $v_{\rm shock}$ is the shock propagation speed
\cite{1992ApJ...393..742E}. This can be written as a distance of about
30 times the mean free path of a photon scattering off electrons and
it is therefore proportional to the electron density.

In RSGs, the shock is expected to travel for about a day until it
reaches an optical depth from infinity small enough for the shock to
emerge. For example, a shock traveling at 10,000 km/s would take 19
hours to traverse 1,000 $R_\odot$. The emergence of the shock itself,
or \emph{shock breakout} (SBO, see review in
\cite{2017hsn..book..967W}), should typically last for about an hour
if it happens from the envelope of a RSG, a timescale which is
determined by the width of the shock precursor (proportional to
density), the shock velocity and some form factors. We call this
scenario an \emph{envelope SBO}, e.g. \cite{2011ApJS..193...20T}.

If a dense wind or atmosphere is present, the ejecta's photosphere at
shock emergence can extend beyond the typical RSG sizes predicted by
stellar evolution theory. This can push shock emergence radially
outwards into lower electron densities, where the shock's radiative
precursor will be significantly more extended for the same optical
depth and will therefore have a longer--lived emergence. Thus, apart
from powering the early light curve via conversion of kinetic energy
into radiation, this dense circumstellar material (CSM) can delay the
shock emergence and extend its duration to several days due to both
the lower electron density and light travel time effects, replacing
the much shorter--lived envelope SBO. We call this scenario a
\emph{wind SBO} \cite{2010ApJ...724.1396O, 2015ApJ...804...28G,
  2017MNRAS.469L.108M, 2018MNRAS.476.2840M, 2017NatPh..13..510Y} or
\emph{extended atmosphere SBO} \cite{2017A&A...605A..83D}.

After shock emergence, the envelope of RSGs should adiabatically cool
until recombination of hydrogen starts. As the RSG envelope expands
and cools, its typical temperature enters the optical range in a
timescale of several days, making SNe II optical light curves (LCs)
rise up to maximum light with a timescale of about 7 days
\cite{2015MNRAS.451.2212G}, with possible sub--classes of relatively
slow and fast rising SNe II \cite{2016ApJ...828..111R}. This timescale
has been used to derive typical RSG stellar radii using analytic
approximations or hydrodynamical models. The observed rise timescales
are shorter than expected and have led researchers to conclude that
either the radii of RSGs are much smaller than predicted by stellar
evolution theory \cite{2015MNRAS.451.2212G, 2017ApJ...848....8R}, or
that perhaps a dense wind SBO is responsible for the fast rise. The
latter is supported by the early spectroscopic detection of narrow
optical emission lines with broad electron--scattering wings in
SN2013fs \cite{2017NatPh..13..510Y}, which are suggestive of slowly
moving shocked material \cite{2017A&A...605A..83D}; and by a recent
analysis of type II SN light curves around maximum light
\cite{2017ApJ...838...28M}. Thus, it is important to understand the
innermost regions of the wind or extended atmospheres of RSGs.

There have been several observational efforts to discover RSG envelope
SBOs using high cadence observations from space
\cite{2008Sci...321..223S, 2008ApJ...683L.131G, 2016ApJ...820...23G,
  2016ApJ...820...57G} and ground--based observatories
\cite{2016ApJ...819....5T} from the UV to optical wavelengths, but
only marginal detections have been achieved (a notable exception was
the serendipitous discovery of the envelope stripped SN IIb 2016gkg
\cite{2018Natur.554..497B}). HiTS (F16) is a survey which uses the
Dark Energy Camera (DECam) to explore the transient sky in timescales
from hours to weeks, monitoring the hour timescale transient sky
during three observational campaigns in 2013, 2014 and 2015.

\section*{Observations and data processing} \label{sec:obs}

The observational strategy of HiTS is described in F16. In 2013 we
surveyed 120 deg$^2$ in the $u$--band with a cadence of 2 hours during
4 consecutive nights. In 2014 and 2015 we surveyed 120 and 150
deg$^2$, with a cadence of 2 and 1.6 hours during 5 and 6 consecutive
nights, respectively. In 2015 the high cadence phase was followed by a
few observations days later, mostly in $g$, but also in $r$ band. The
2014 and 2015 data were analyzed in real--time, with candidates being
generated and filtered only minutes after the end of every
exposure. Thanks to these real--time capabilities, in 2014 and 2015 we
were able to trigger spectroscopic follow--up observations for a few
of the most nearby candidates, with hour timescale spectroscopic
follow up capabilities in 2015 using the Very Large Telescope. No
clear signatures of envelope SBOs were found and the very fast
spectroscopic observations were never triggered, but we obtained
spectra for 18 objects after the main high cadence phase of the
observations for classification purposes: using NTT (provided by the
PESSTO collaboration \cite{2015A&A...579A..40S}), SOAR and VLT (see
Methods). These spectra were used for direct classification, but also
for testing a LC based classifier. 11 SN candidates were
spectroscopically classified as SNe Ia and 7 SN candidates were
classified as either SNe II or showed a blue continuum consistent with
SNe II.

The DECam data processing from raw image to light curve creation is
discussed in Methods. Although the resulting LCs did not show the
signature of envelope SBOs it was evident that the first few days of
evolution of synthetic light curves which do not include CSM, and
which show an hour timescale SBO signature, did not resemble our
observations. In order to investigate this further we used models from
\cite{2017MNRAS.469L.108M, 2018MNRAS.476.2840M} (hereafter M18),
which include CSM, and developed tools that can be used to aid the
classification and physical interpretation of these LCs.

\paragraph{Light curve based classification} \label{sec:classifier}

\begin{figure*}[!ht]
  \centering
  \includegraphics[width=0.8\hsize, bb=0 0 1008 720]{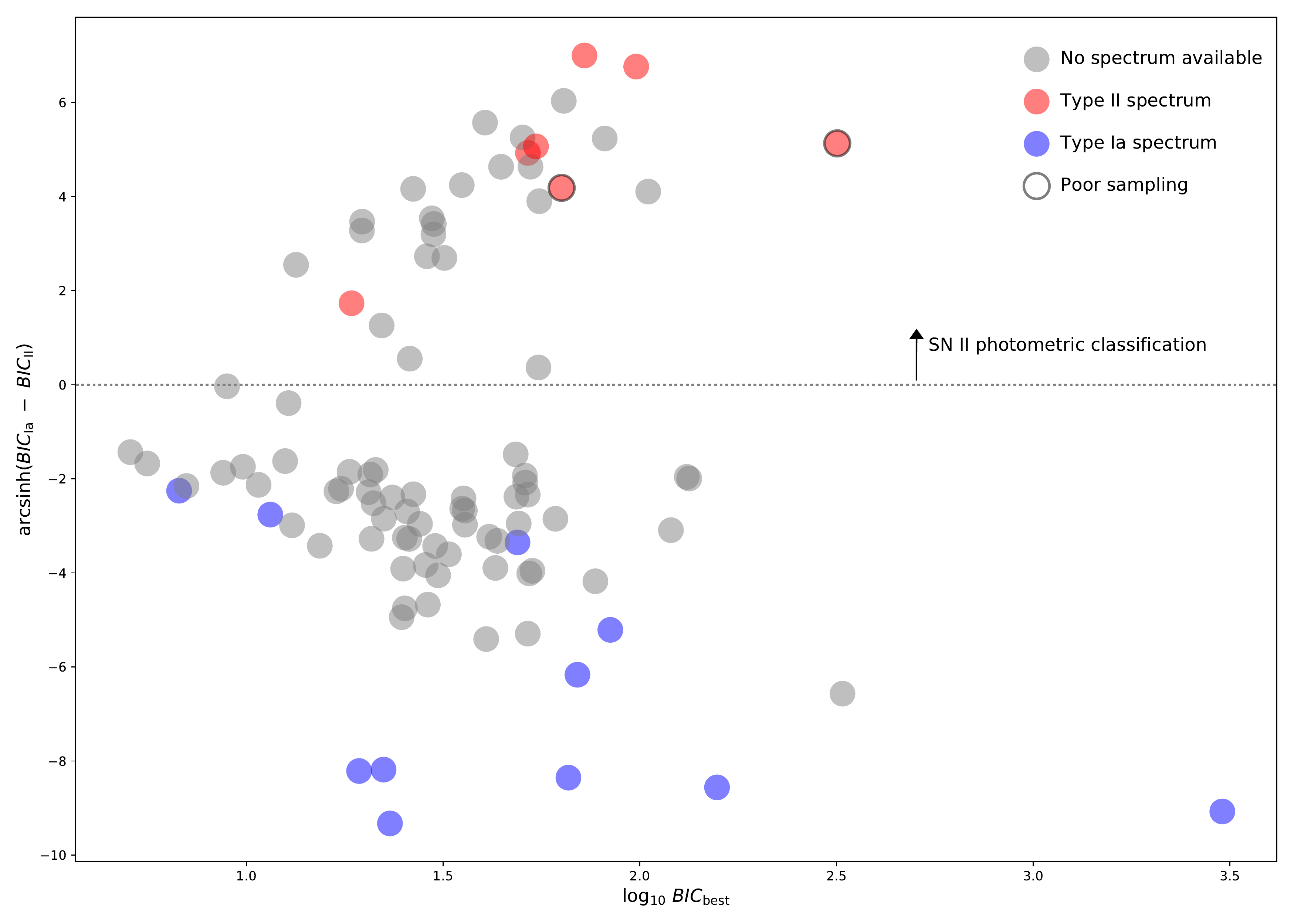}
  \caption{Light curve based classification implemented in this
    work. We show the arcsinh of the BIC differences between our SN II
    and SN I models as discussed in the text. Note that all SNe with
    spectroscopic information were correctly classified. There were
    two SNe with spectra which were not included in the analysis
    because of their poor sampling at rise.}
  \label{fig:classification}
\end{figure*}

When no spectral information could be acquired we used the SN early
LCs for classification. Since our LCs are based on image differences,
which sometimes contain a very recent template, our observables are LC
differences between two points in time, i.e. we take into account that
our templates can contain some SN flux. Comparing these observables
against model predictions for a class of events we can perform model
selection and classification. For the RSG SNe II we use the family of
models from M18. The most important remaining classes are
thermonuclear SNe (Ia) and envelope--stripped core collapse SNe
(Ib/c), which in most cases are explosions whose rise to maximum is
dominated by the deposition of energy from $^{56}$Ni starting from a
compact configuration \cite{2016MNRAS.457..328L}. Therefore, as a
simple approximation we use SN Ia spectral templates from
\cite{2007ApJ...663.1187H} and allow for a broad range of stretch and
scale factors to account for greater diversity. Then, using a Markov
Chain Monte Carlo (MCMC) sampler (as explained in Methods) we compute
a median log--likelihood for these two families of models and select
between them using the Bayesian Information Criterion (BIC). With this
method we correctly classify all 18 SNe with spectroscopic
classification: 7 SNe II and 11 SNe Ia (see
Figure~\ref{fig:classification}), which highlights the power of using
early time photometry for classification.

We removed SNe which had a maximum time span of less than one week,
which eliminated all candidates from 2013A and one SN with spectra
from 2014A. Also, since we are mostly interested in the effect of mass
loss and wind acceleration, which affect mostly the initial rise after
emergence, we remove those SNe with a poor sampling at rise. We define
a poor sampling at rise as those cases where three or more continuous
days during rise have no data. This includes SNe which have gaps in
the data of two or more consecutive nights during the initial rise (5
SNe discovered at the end of the 15A campaign, which had a logarithmic
cadence); and SNe where we cannot rule out that they were seen only
three or more days after their first light. For the latter case we
remove those SNe that according to our posterior distribution had more
than 10\% probability of having been observed only three or more days
after first light, defining the first light as the moment when the
absolute magnitude in $g$ band reaches -13 mag. This additional filter
resulted in the removal of two SNe.

\begin{figure*}[!ht]
\includegraphics[width=0.9\hsize, bb=0 0 800 800]{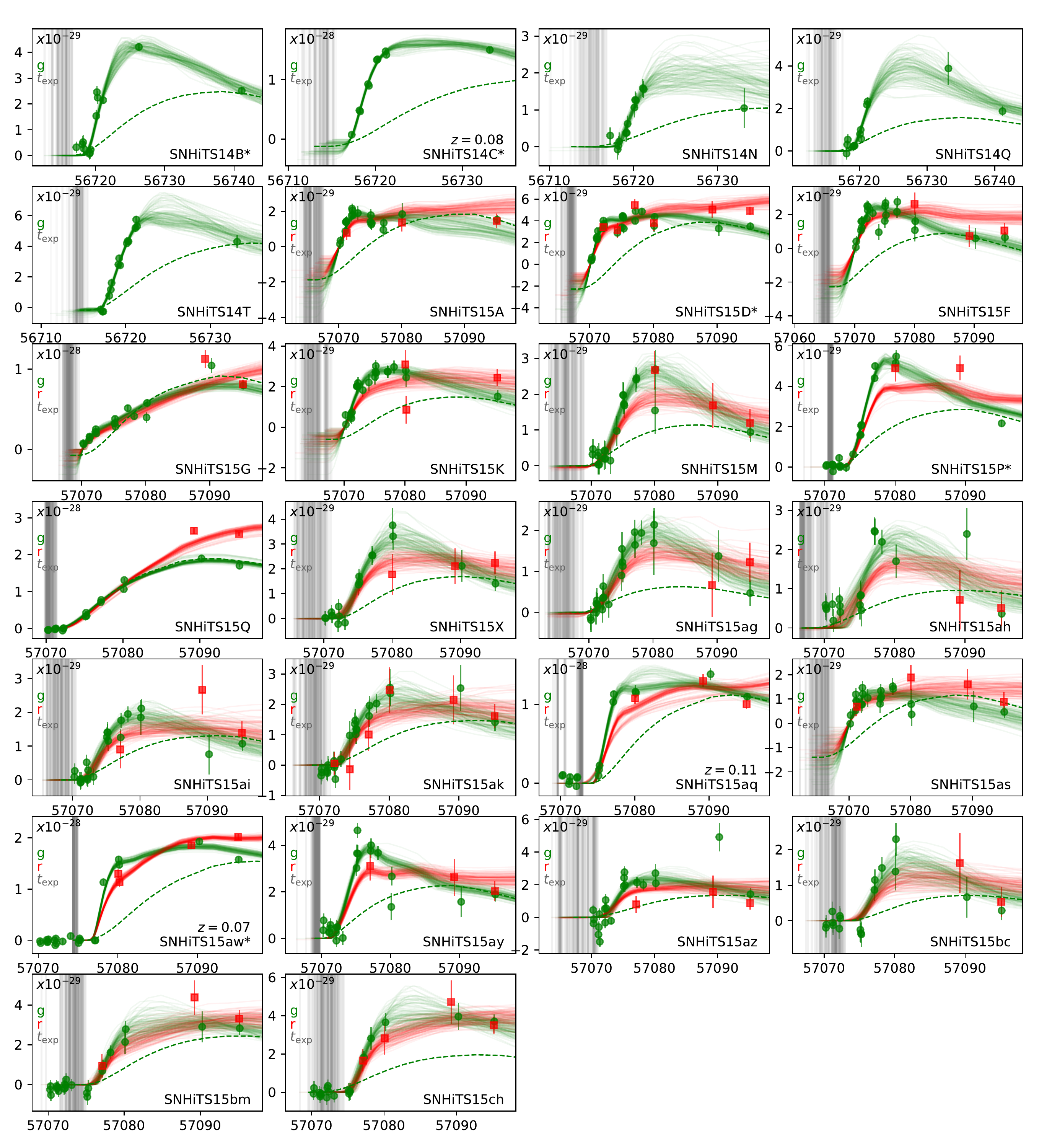}
\caption{Observation minus template flux in [erg/s/cm$^2$/Hz] vs MJD
  time for the Type II SN candidates in $g$ (green circles) and $r$
  (red squares) bands with 1--$\sigma$ error bars. Flux differences
  from 100 models randomly sampled from the posterior are shown
  (continuous lines with the same colors per band), including one $g$
  band example with mass loss rate set to zero (green dashed
  lines). Explosion times are indicated by the vertical gray
  lines. SNe with spectroscopic classification are shown with an
  asterisk, and spectroscopic host galaxy redshifts are indicated when
  available. Most SNe show a steep initial rise, the signature of the
  wind shock breakout.}
\label{fig:LCs}
\end{figure*}

\paragraph*{Early SN II light curves and models}
The resulting sample of 26 early time SN II candidate LCs discovered
by the HiTS survey are shown in Figure~\ref{fig:LCs}. There is no
similar sample of SN II early LCs in the literature in terms of
cadence, i.e. every 2 or 1.6 hours during the night for several
consecutive nights, and size. We will include five additional SNe II
from the literature in the analysis, all with well sampled optical
early light curves: SN2006bp \cite{2007ApJ...666.1093Q}, PS1-13arp
\cite{2015ApJ...804...28G}, SN2013fs \cite{2017NatPh..13..510Y},
KSN2011a and KSN2011d \cite{2016ApJ...820...23G}. The SN II candidates
show a fast rise with a timescale of a few days which is suggestive of
the wind / extended atmosphere SBO scenario \cite{2015ApJ...804...28G,
  2017MNRAS.469L.108M, 2017A&A...605A..83D}. We also show for
comparison 100 LCs sampled from the posterior distribution of physical
parameters using a grid of SN II explosion models from M18.  Briefly,
we sample the posterior distribution of physical parameters such as
explosion time, attenuation, redshift, progenitor mass, explosion
energy, mass loss rate ($\dot M$) and wind acceleration parameter
($\beta$), using a MCMC sampler as explained in the Suppplementary
Information. For reference, we also show one $g$ band LC sampled
randomly from the posterior where we have set the mass loss to
zero. It can be seen that for most LCs only models with a significant
mass loss rate match the fast rise observed in the data.

\section*{Synthetic models} \label{sec:models}

Models from M18 assume confined steady--state winds with an extended
acceleration length scale. They form a grid of 518 sets of spectral
time series spanning different main sequence masses and explosion
energies, but also different mass loss properties. The RSG SN
progenitors and the CSM around them are constructed in the same way as
in M18. We provide a short summary of the progenitors and CSM.

The public stellar evolution code \texttt{MESA}
\cite{2011ApJS..192....3P,2013ApJS..208....4P,2015ApJS..220...15P}
is used to compute the RSG SN progenitors of the zero-age
main-sequence (ZAMS) masses of 12, 14, and $16~M_\odot$ with solar
metallicity.  We set the model maximum mass to $16~M_\odot$ because
the maximum mass of SN~II progenitors is estimated to be at around
$16.5~M_\odot$ based on the RSG progenitor detections
\cite{2009MNRAS.395.1409S}.  The progenitors are evolved by using the
Ledoux criterion for convection with a mixing-length parameter of 2.0
and a semiconvection parameter of 0.01. Overshooting is taken into
account on top of the hydrogen-burning convective core with a step
function using the overshoot parameter of $0.3 H_P$, where $H_P$ is
the pressure scale height. We use the `Dutch' mass-loss prescription
in \texttt{MESA} without scaling for both hot and cool stars.  The RSG
progentiors are evolved to the core oxygen burning, from which the
hydrogen-rich envelope structure hardly changes until the core
collapse. The final progenitor properties are summarized in Table~1.

\begin{table} \label{tab:progenitor}
  \begin{centering}
    \begin{tabular}{cccc}
      ZAMS mass & Final mass & Final H-rich envelope mass  & Final radius \\
      \hline
      $12~M_\odot$ & $10.3~M_\odot$ & $6.1~M_\odot$ & $607~R_\odot$ \\
      $14~M_\odot$ & $11.4~M_\odot$ & $6.2~M_\odot$ & $832~R_\odot$ \\
      $16~M_\odot$ & $12.0~M_\odot$ & $5.8~M_\odot$ & $962~R_\odot$ \\
    \end{tabular}
    \caption{RSG progenitor properties. }
  \end{centering}
\end{table}

A CSM structure with density $\rho_\mathrm{CSM}(r) = \dot{M}/(4\pi
v_\mathrm{wind})r^{-2}$ is attached on top of these progenitors.
Here, $\dot{M}$ is the progenitor's mass-loss rate, $r$ is the
distance from the center of the star and $v_\mathrm{wind}$ is the wind
velocity.  We do not change $\dot{M}$ with $r$ in our models. Instead
of $\dot{M}$, we take the radial change of $v_\mathrm{wind}$ caused by
the wind acceleration into account. The simple $\beta$ velocity law
for the wind velocity, i.e.,
\begin{equation}
v_\mathrm{wind} (r) = v_0 + (v_\infty - v_0) \left( 1 - \frac{R_0}{r} \right)^\beta,
\end{equation}
where $v_0$ is the initial wind velocity, $v_\infty$ is the terminal
wind velocity, and $R_0$ is the wind launching radius that we set at
the stellar surface, is adopted. $v_\infty =
10~\mathrm{km~s^{-1}}$ is fixed in our models. $v_0$ is chosen so that the CSM density is
smoothly connected from the surface of the progenitors and it is less
than $10^{-2}~\mathrm{km~s^{-1}}$. We take $\beta$ between 1 and 5 because OB
stars have $\beta\simeq 0.5-1$ \cite{1996A&A...305..171P} and
RSGs are known to experience slower wind acceleration than OB stars,
i.e., $\beta > 1$
e.g. \cite{2010ASPC..425..181B,2004MNRAS.355.1348M}. For instance, a
RSG $\zeta$ Aurigae is known to have $\beta\simeq 3.5$
\cite{1996ApJ...466..979B}.

\begin{figure*}[!ht]
  \vbox{
    \hbox{
      \hspace{1cm}
      \includegraphics[width=0.25\hsize, bb=0 0 350 600]{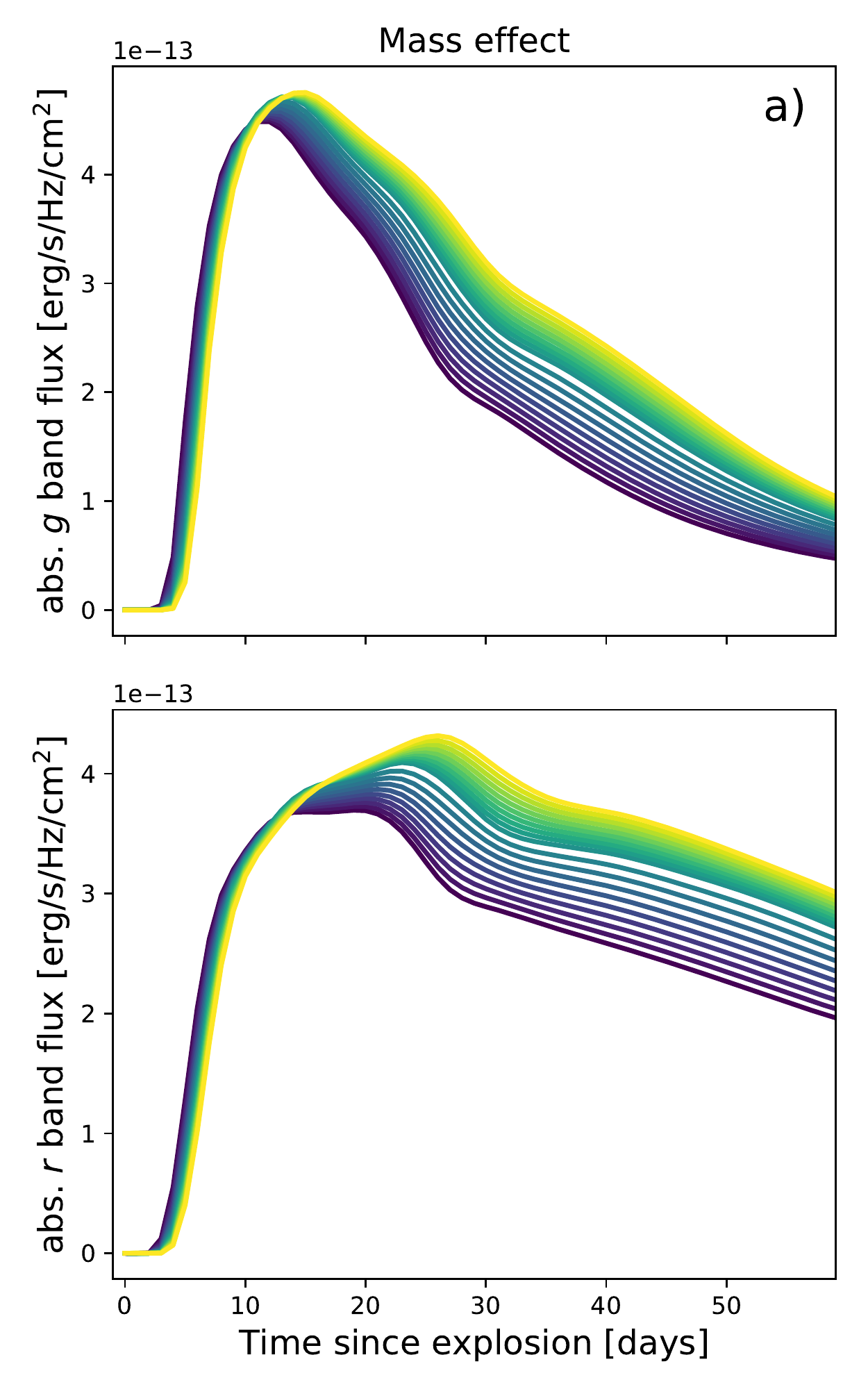}
      \includegraphics[width=0.25\hsize, bb=0 0 350 600]{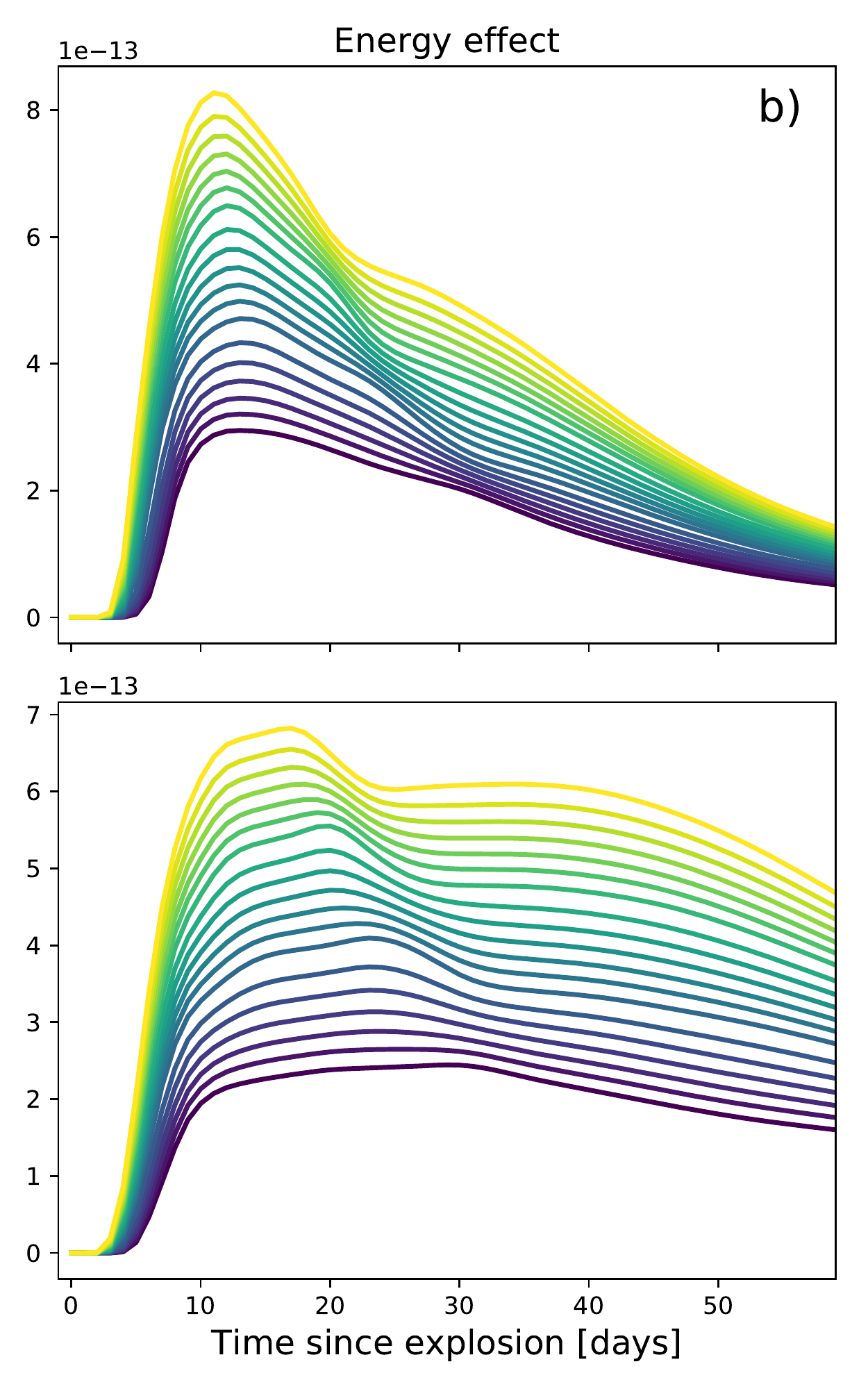}
      \includegraphics[width=0.25\hsize, bb=0 0 350 600]{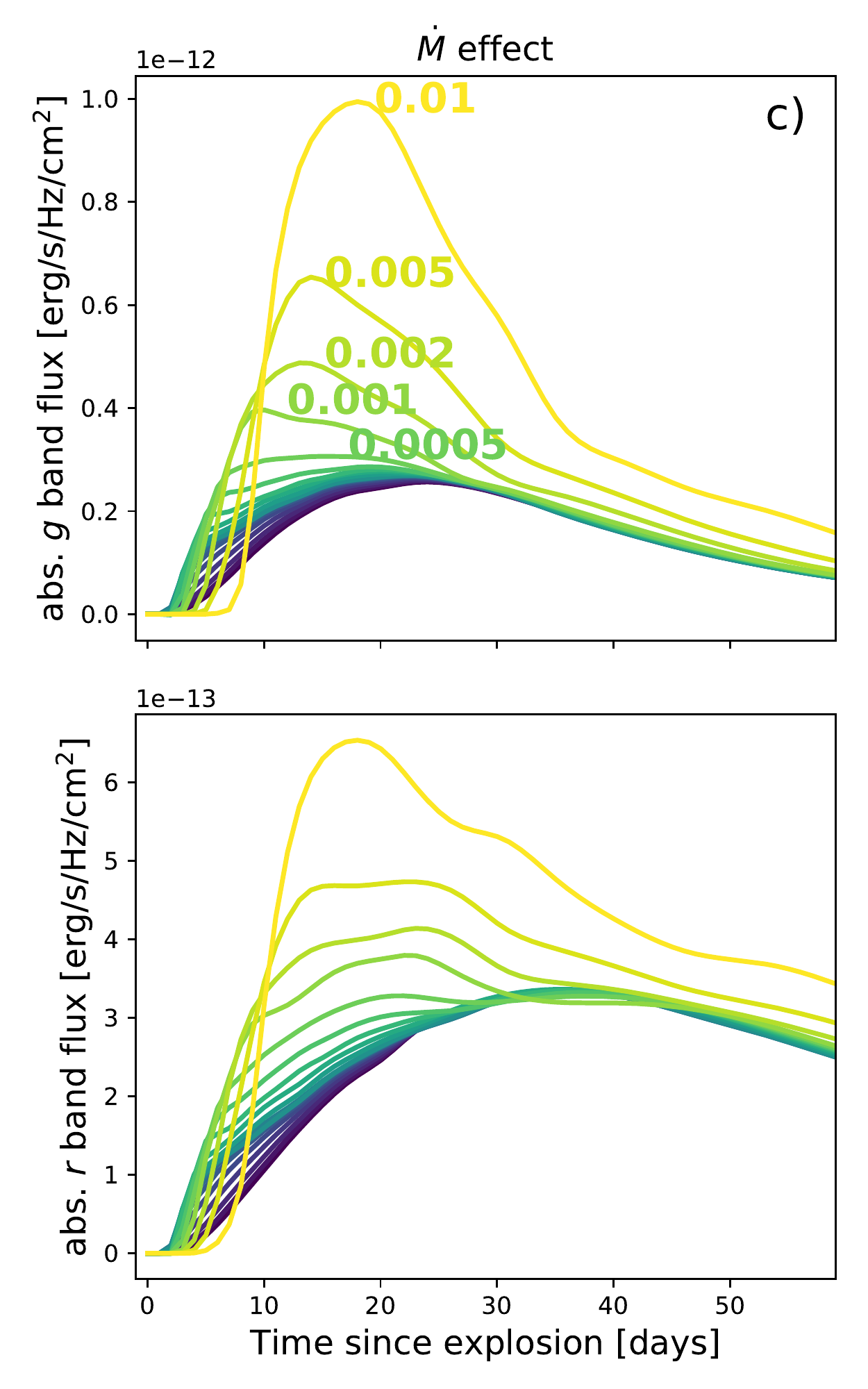}
    }
    \hbox{
      \hspace{1cm}
      \includegraphics[width=0.25\hsize, bb=0 0 350 600]{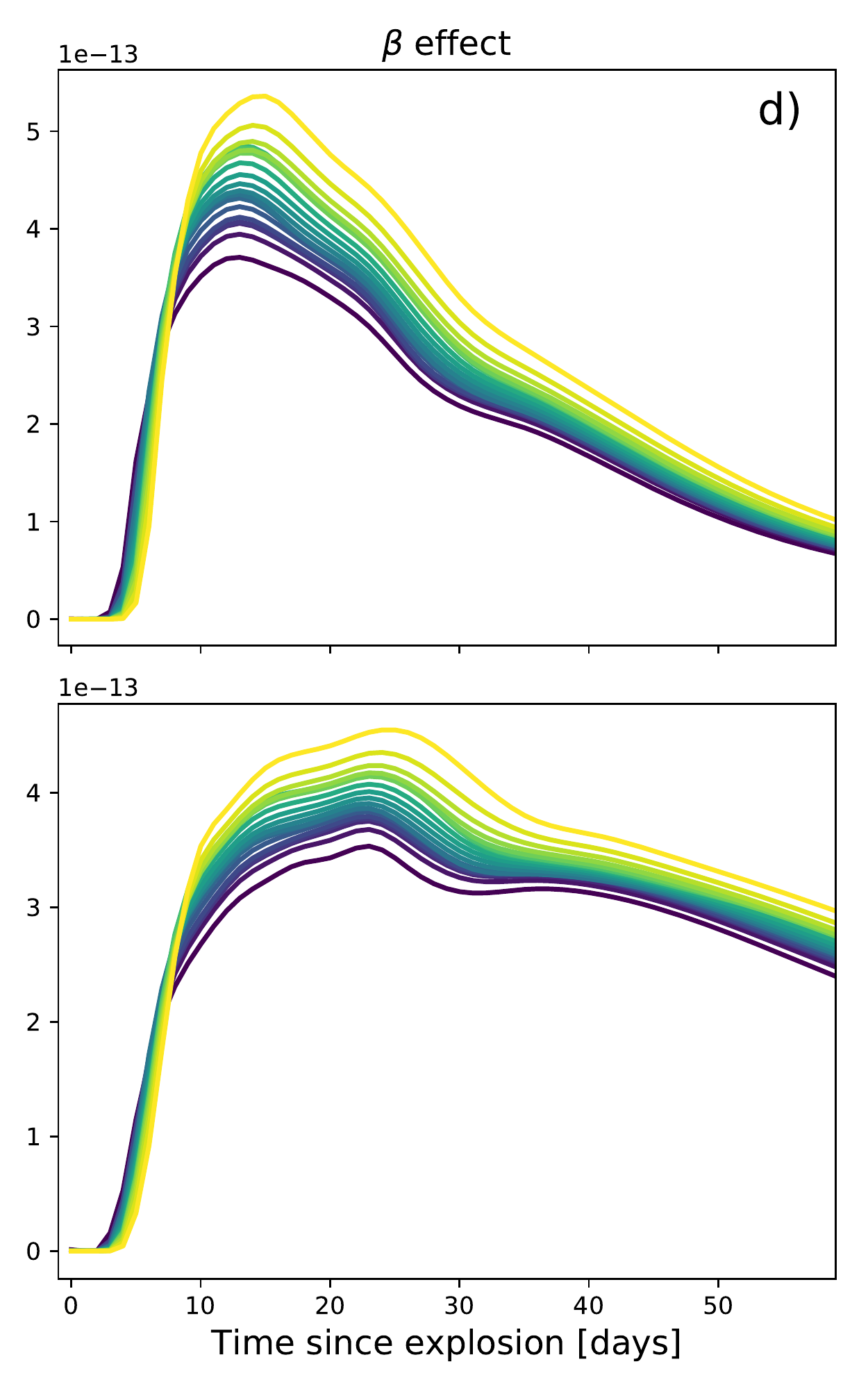}
      \includegraphics[width=0.25\hsize, bb=0 0 350 600]{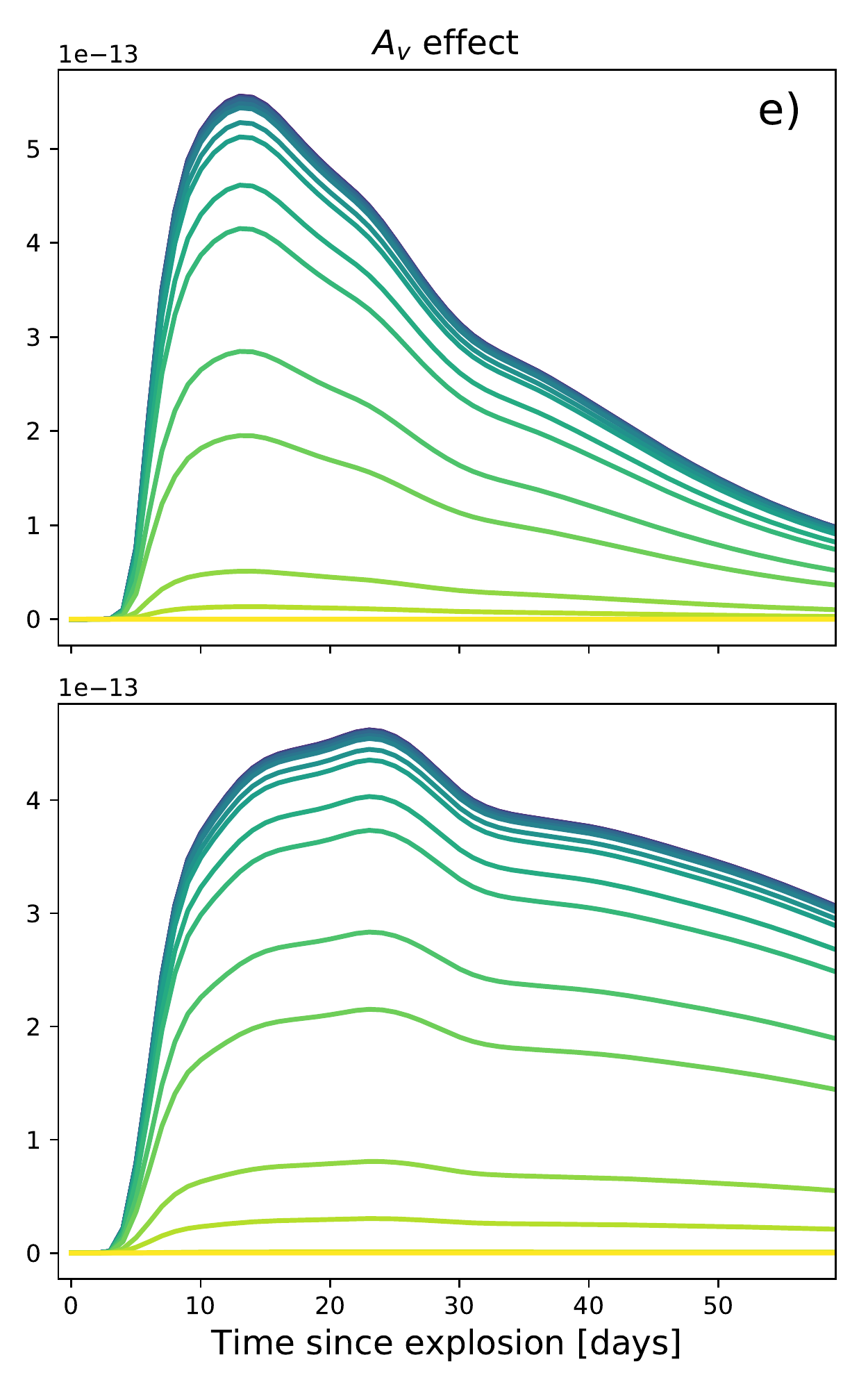}
      \includegraphics[width=0.25\hsize, bb=0 0 350 600]{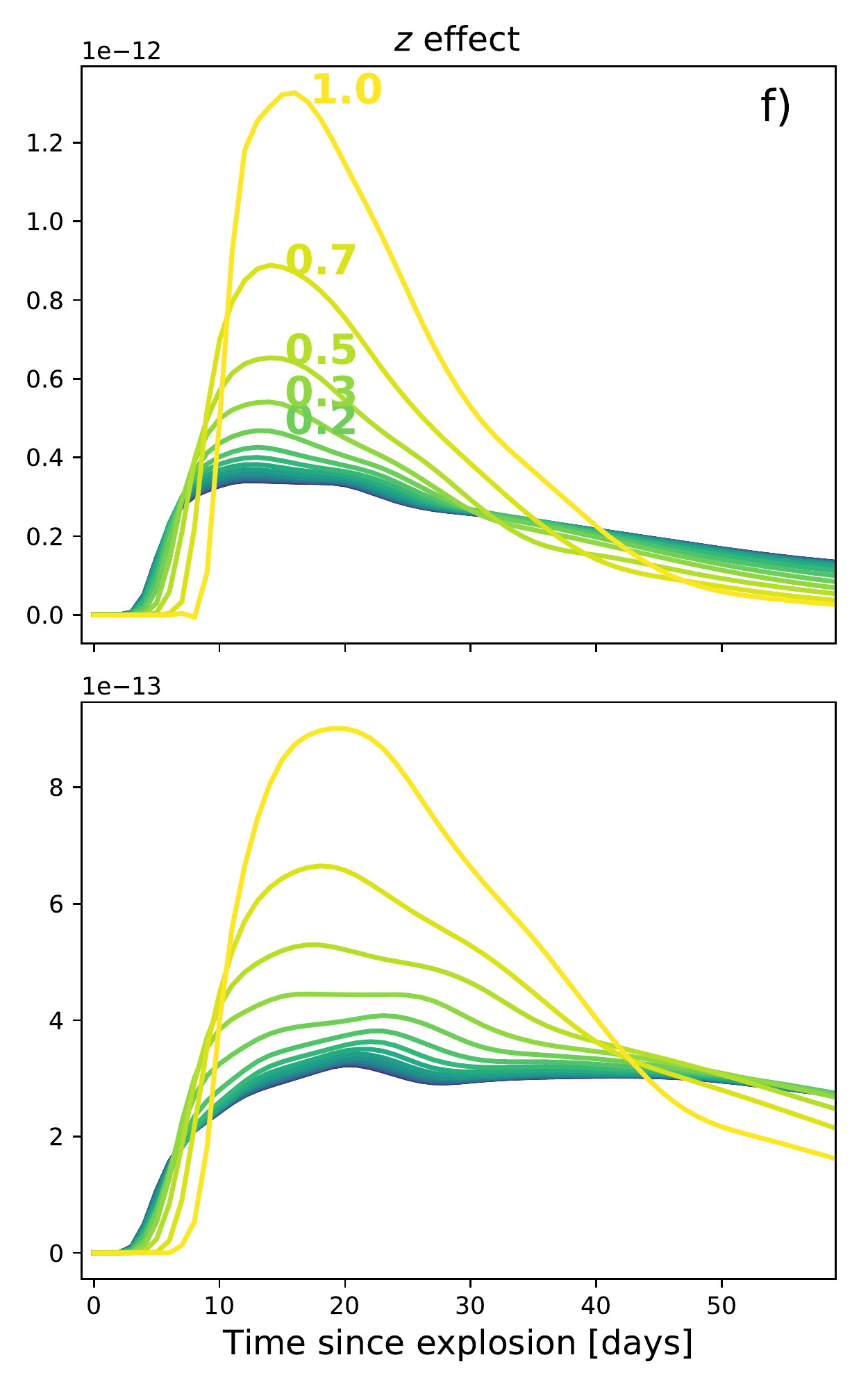}
    }
  }
  \caption{Interpolated synthetic absolute fluxes (obtained from the
    predicted apparent magnitude minus the distance modulus) in the
    $g$ and $r$ bands after varying the progenitor mass linearly
    between 12 and 16 $M_\odot$ (a), the energy linearly between 0.5
    and 2 B (b), $\dot M$ logarithmically between $10^{-8}$ and
    $10^{-2}$ \msunyr\ (c), $\beta$ linearly between 1 and 5 (d),
    $A_V$ logarithmically between $10^{-4}$ and $10$, and $z$
    logarithmically between 0.001 and 1. $\dot M = 0$ is arbitrarily
    associated to $10^{-8}$ \msunyr\ in order to perform the
    logarithmic interpolation. Values range from blue to yellow in an
    increasing array of 19 values for each parameter being varied
    (selected values shown for $\dot M$ and $z$). The parameters not
    being varied are fixed at a mass of 14 $M_\odot$, energy of 1 B,
    $\dot M$ of $5 \times 10^{-3}$ \msunyr, $\beta$ of 3.5, $A_V$ of
    0.1, and redshift of 0.22.}
  \label{fig:interp_mdot}
\end{figure*}

RSG mass loss rates are known to be dependent on the luminosity of the
progenitor star, but estimations differ greatly in the literature. For
the typical luminosities observed in RSG SN II progenitors, between
$10^4$ an $10^5$ $L_\odot$ \cite{2009MNRAS.395.1409S}, derived mass
mass loss rates range from between $10^{-7}$ and $10^{-6}$ \ \msunyr
\ \cite{2011A&A...526A.156M} to between $10^{-5}$ and $10^{-4}$
\ \msunyr \ \cite{2017MNRAS.465..403G}. The M18 models assume a wider
range of mass loss rates, including zero mass loss and a range of
values between $10^{-5}$ and $10^{-2}$ \msunyr, always assuming
\emph{enhanced} mass loss episodes before explosion extending up to
$10^{15}$ cm from the progenitor. Note that a shock travelling at
10,000 km/s would take about 11 days to cover $10^{15}$ cm. Recent
optical and X-ray modeling of the SN IIP 2013ej suggests that the
dense CSM radius may be relatively small ($10^{14}$ cm), although
there are some model uncertainties \cite{2018arXiv180407312M}.

\paragraph{Hydrodynamics and radiative transfer}

The LCs from the explosions of the above mentioned progenitors with
CSM are numerically obtained by using the one-dimensional multi-group
radiation hydrodynamics code \texttt{STELLA}
\cite{1998ApJ...496..454B, 2000ApJ...532.1132B,
  2006A&A...453..229B}. \texttt{STELLA} follows the evolution of the
spectral energy distributions (SEDs) at each time step and the
multi-color LCs from the explosions are obtained by convolving the
filter functions to the numerical SEDs. We use the standard setup for
the SED resolution, i.e., 100 bins that are distributed between
1~\AA\ and 50000~\AA\ in a log scale. The code initiates the
explosions as thermal bombs. All the models have $0.1~M_\odot$ of the
radioactive $^{56}$Ni at the center but it does not affect the early
LCs we are interested in. Every progenitor model with CSM is exploded
with four different explosion energies to investigate the dependence
on the explosion energy; $5\times 10^{50}$~erg, $10^{51}$~erg,
$1.5\times 10^{51}$~erg, and $2\times 10^{51}$~erg. The effects of
mass, energy, mass loss rate, wind acceleration parameter, attenuation
and redshift on the optical light curves can be seen in
Figure~\ref{fig:interp_mdot}.

\section*{Results}

\begin{figure*}[!ht]
\includegraphics[width=\hsize, bb=0 0 850 550]{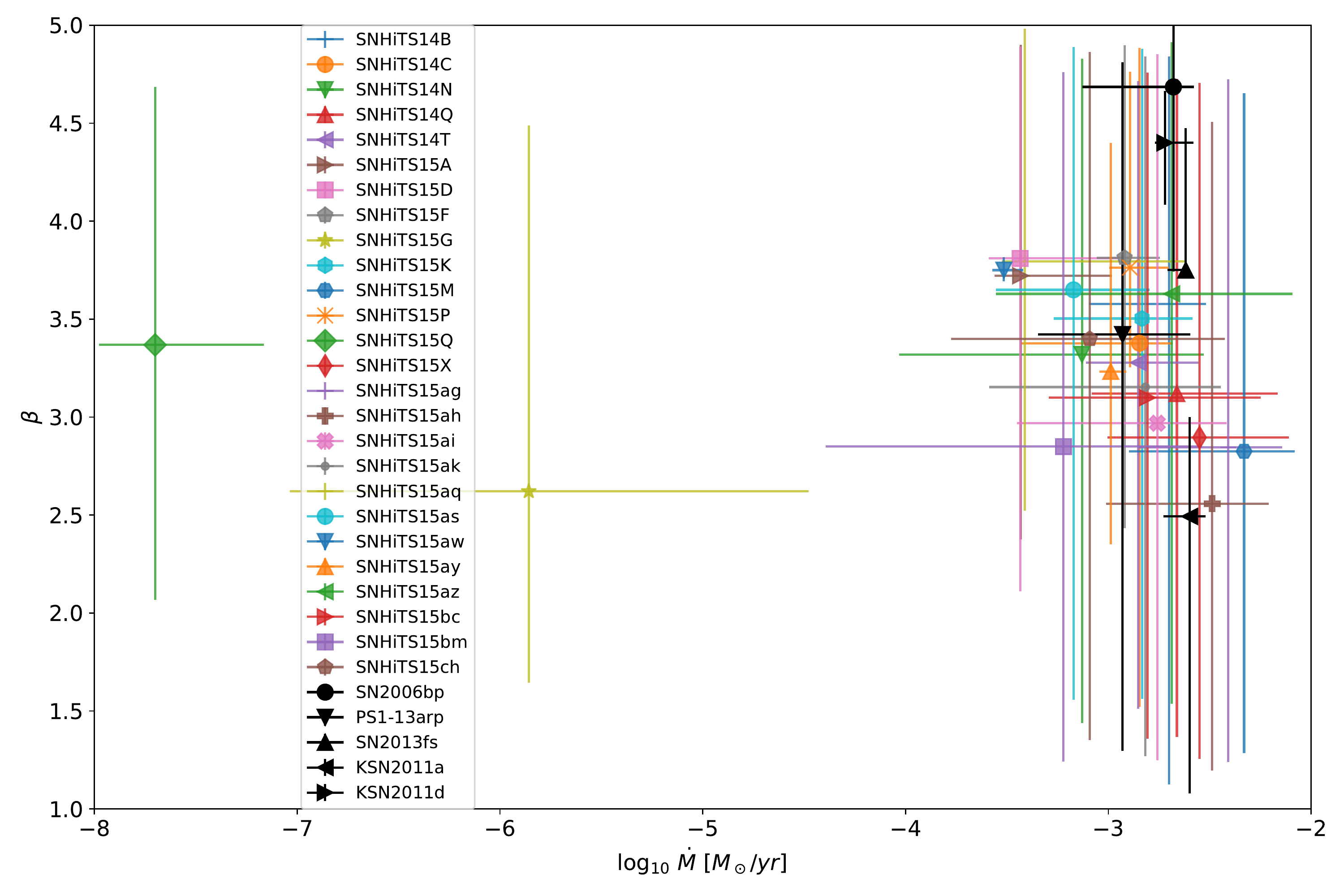}
\caption{Distribution of inferred $\beta$ and $\dot M$ values for the
  sample of 26 early SNe II candidates from HiTS, as well as selected
  SNe II from the literature with well sampled optical early light
  curves (see Figure~\ref{fig:MCMCother}). The error bars correspond
  to the percentiles 5, 50 and 95 of the marginalized posterior
  distributions in $\beta$ and $\dot M$. Note that the available
  models range from no mass loss, which we arbitrarily assign
  10$^{-8}$ \msunyr, to larger values in a grid between $10^{-5}$ and
  $10^{-2}$ \msunyr. Models between $10^{-8}$ and $10^{-5}$
  \msunyr\ are logarithmically interpolated between the zero and
  $10^{-5}$ \msunyr\ mass loss models.}
\label{fig:betavsmdot}
\end{figure*}

Early time SN II optical LCs are affected by many physical parameters
(see Figure~\ref{fig:interp_mdot}). These can change the light curve
normalization, its characteristic rise time and general shape, the
ratio between bands or a combination of them. The parameters which
affect the normalization the most are first redshift and $A_V$, and
then mass loss rate and energy (factor of 10 difference in peak
apparent $g$ mag range for the full grid of values tested). The
parameters which affect the light curve shape the most are the mass
loss rate and redshift, with somewhat similar effects, but very
different normalizations which help break their degeneracy. Mass and
$\beta$ also affect the normalization, light curve shape and color,
but not as strongly as the other variables.

In this work we have concentrated on those variables constraining the
CSM density profile, which in the M18 models used in this work is
parametrized via two numbers: $\dot M$ and $\beta$, assuming a
terminal wind velocity of 10 km/s and and a maximum CSM radius of
$10^{15}$ cm. Therefore, we focus the discussion on these last two
quantities.

\begin{figure*}[!ht]
  \begin{center}
    \hspace{-2cm}
  \includegraphics[width=0.7\hsize, bb=0 0 700 700]{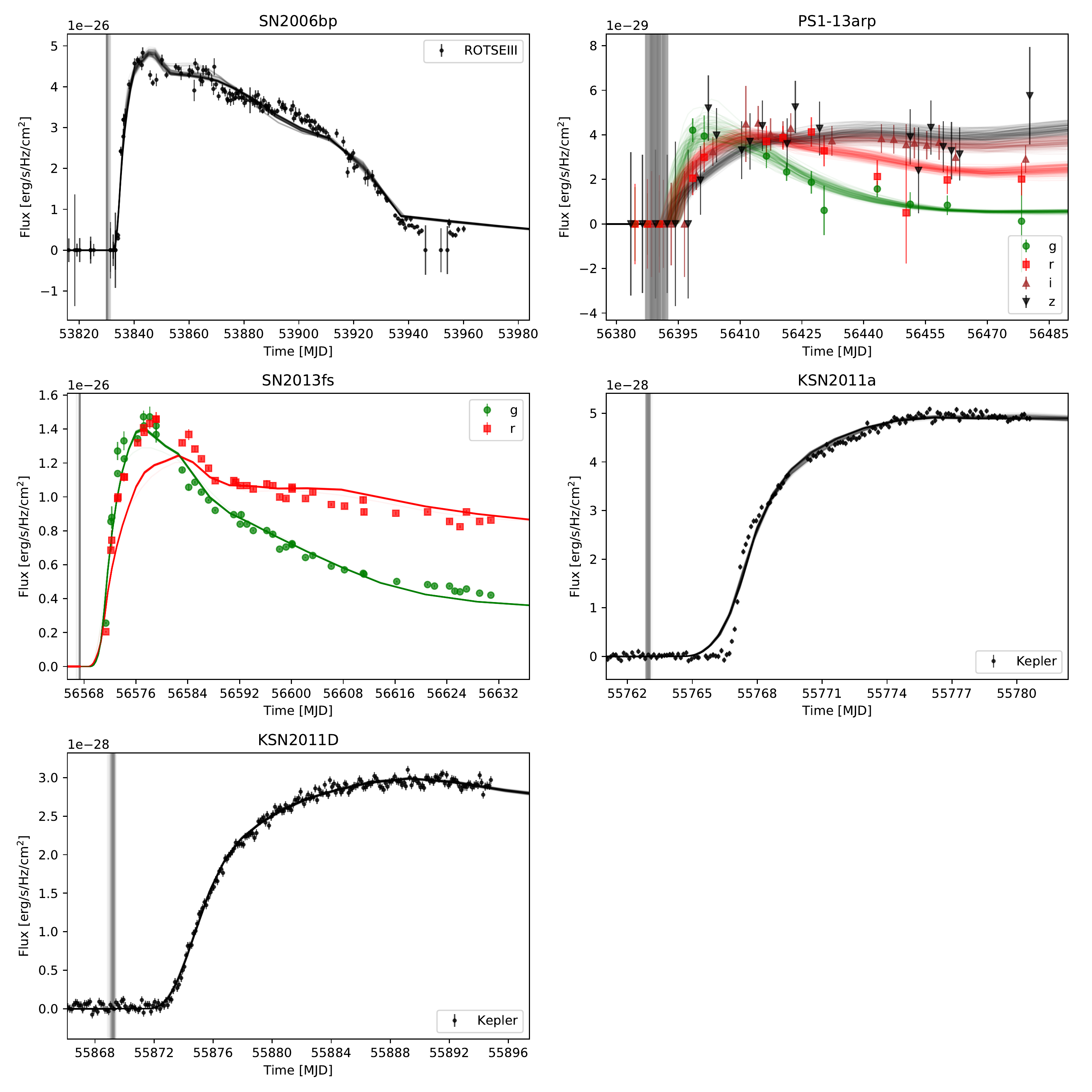}
  \caption{Similar to Figure~\ref{fig:interp_mdot}, but for selected
    SNe II from the literature with well sampled optical early light
    curves: SN2006bp \cite{2007ApJ...666.1093Q}; PS1-13arp
    \cite{2015ApJ...804...28G}; SN2013fs \cite{2017NatPh..13..510Y};
    KSN2011a and KSN2011d \cite{2016ApJ...820...23G} in the indicated
    bands. The 5, 50 and 95 percentiles for the mass loss and wind
    acceleration marginalized posterior distributions are shown in
    Figure~\ref{fig:betavsmdot}.}
  \end{center}
  \label{fig:MCMCother}
\end{figure*}

The posterior distributions of $\dot M$ and $\beta$ for the entire
sample of HiTS SN II candidates, as well as five additional SNe II
from the literature with well sampled optical early light curves are
summarized in Figure~\ref{fig:betavsmdot}. In the $x$ and $y$ axis we
show the 5, 50 and 95 percentiles of the marginalized posterior
distributions over all variables except $\dot M$ or $\beta$,
respectively (error bars cover the percentiles 5 to 95, and they
intersect at percentiles 50). Mass loss rates smaller than $\dot M =
10^{-4}$ \msunyr\ are allowed in only two out of 26 HiTS SNe II
candidates, and in none of the five SNe II from the literature. We
also constrain $\beta$ within the range of values covered by our grid
of models ($1 < \beta < 5$).  Although we find significantly large
values of $\beta$ for some SNe, we do not find a significant excess of
large values within the interval $1 < \beta < 5$ in the sample. The
literature comparison light curves and their posterior sampled
synthetic light curves are shown in Figure~\ref{fig:MCMCother}. The
complete list of inferred parameters are shown in Supplementary Tables
1 and 2.

We have found evidence for a density profile consistent with large
mass loss rates $\dot M > 10^{-4}$ \msunyr\ in 29 out of 31 SNe II and
evidence for a density profile consistent with relatively slow wind
acceleration (large $\beta$) for some SNe. These constraints suggest
that CSM densities in the vicinity of RSGs before explosion are larger
than $10^{-13}$ g cm$^{-3}$ at $10^{14}$ cm from the center of the
star (about 1400 $R_\odot$). Assuming an electron scattering opacity
consistent with fully ionized CSM made of hydrogen and helium at solar
abundance ($\kappa\ =\ 0.34\ {\rm cm}^2 ~{\rm g}^{-1}$), the optical
depth $\tau \approx 1$ would occur at typical radii between $10^{14}$
and $10^{15}$ cm and at typical densities between $10^{-12}$ and
$10^{-14}$ g cm$^{-3}$. The typical densities implied from this
analysis are in very good agreement with those derived
spectroscopically for SN2013fs in \cite{2017NatPh..13..510Y}, and the
mass loss rates are consistent with those derived using NUV data for
PS1-13arp.

The large mass loss rates derived in this work imply total
CSM masses typically between 0.1 and 1.0 $M_\odot$, and typically
between 0.01 and 0.1 $M_\odot$ above the breakout radius ($\tau \le
c/v$).  These mass loss rates could not be sustained for large periods
of time before explosion as noted by \cite{2017NatPh..13..510Y},
otherwise we would see Type IIn--like features at late times (narrow
optical emission lines with electron scattering wings) and the
progenitor star would lose its H--rich envelope before explosion. This
favors the interpretation of dense atmospheres or confined accelerated
wind SBOs. However, it is interesting to note that the large derived
mass loss rates appear to match those from the Type IIb SN2013cu
\cite{2014Natur.509..471G, 2014A&A...572L..11G, 2016MNRAS.455..112G},
a likely post--RSG star \cite{2016MNRAS.455..112G}. The fact that most
of this CSM would be optically thick at breakout implies that apart
from delaying the SBO, most of the shocked material would contribute
to the optical light curve via its cooling after breakout. In fact,
the differences between LCs with varying CSM radii in
\cite{2018MNRAS.476.2840M} are in part explained by this effect.

The range of wind acceleration parameter $\beta$ favored by these
observations is consistent with those derived for many normal RSGs
\cite{1985A&A...147..103S}. However, models assume enhanced mass loss
before explosion and our observations cannot be used to discard
whether such enhanced mass loss is required to produce the high
density CSM being shocked. The large mass loss rates suggest that
either RSGs undergo this enhanced mass loss before explosion or that
we are probing complex RSG atmospheres and not an accelerated
wind. The fact that high resolution observations of nearby RSGs show
complex CSM environments \cite{2017Natur.548..310O} suggests that
large CSM densities may in fact be possible without requiring
accelerated or enhanced mass loss.

An important implication from this work is that if high CSM densities
are present in most SNe II, the hour timescale, high density RSG
envelope SBOs may be rare in nature. With the exception of two slow
rising SNe out of 26 SN II candidates, large CSM densities were always
inferred (see Methods for the significance of this result). The
photons produced when the shock crosses the outer envelope of these
stars and enters the high density CSM would not be able to diffuse in
a timescale of hours, but would instead take days to emerge in the
wind SBO. Only if the CSM radii are much smaller than the values
tested in this work could a short timescale signature be observed (see
Figure~7 in \cite{2018MNRAS.476.2840M}). Given that we have found
breakout densities to be similar for different CSM density profiles,
measuring the lag between the time of explosion (via neutrinos or
gravitational waves in a nearby SNe II) and shock emergence (from one
to 12 days after explosion in the parameter space probed by M18),
could greatly help understand the true extent of the CSM.

These results also suggest that pre--explosion radii cannot be derived
using the early LC of SNe II, at least at the optical bands, because
the rise to maximum will not be dominated by the adiabatic cooling of
the envelope, but by the wind SBO instead. Furthermore, we have shown
that it is possible to accurately separate compact (SNe I) from
extended (SNe II) explosions using early time information, opening the
possibility of using high cadence observations as a tool for detection
and classification of SNe, including potentially distant SNe II which
can be used for cosmological applications, where precise explosion
time determinations are required \cite{2014AJ....148..107R}.

Finally, we conclude that it may be difficult for future high cadence
surveys such as ZTF to systematically discover RSG envelope
SBOs. Instead, they will be able to increase the sample of wind SBO
events significantly. This also implies that the deep drilling fields
from LSST will be an important source of wind SBO events at all
redshifts, allowing for the study of the CSM around RSG stars with
cosmic age. We expect that the analysis of the physical parameters
inferred from populations of events, such as the one presented in this
work, will become part of the standard set of tools for both the
classification and characterization of large volumes of transients in
the future.

All correspondence and requests for material in relation to this work
should be sent to Francisco F\"orster
(\verb+francisco.forster@gmail.com+).

\section*{Acknowledgments}

We thank Luc Dessart, Keiichi Maeda and Karim Pichara for useful
discussions. F.F. and T.J.M. thank the Yukawa Institute for
Theoretical Physics at Kyoto University, where part of this work was
initiated during the YITP-T-16-05 on ``Transient Universe in the Big
Survey Era: Understanding the Nature of Astrophysical Explosive
Phenomena''. T.J.M. is supported by the Grants-in-Aid for Scientific
Research of the Japan Society for the Promotion of Science (16H07413,
17H02864). Powered@NLHPC: this research was partially supported by the
supercomputing infrastructure of the NLHPC (ECM-02). Numerical
computations were partially carried out on PC cluster at Center for
Computational Astrophysics, National Astronomical Observatory of
Japan. We acknowledge support from Conicyt through the infrastructure
Quimal project No. 140003. F.F., J.S.M, E.V. and S.G. acknowledge
support from Conicyt Basal fund AFB170001. F.F., G.C-V., A.C., P.E.,
M.H., P.H., H.K., J.M, G.M, F.O., G.P., A.R. and I.R. acknowledge
support from the Ministry of Economy, Development, and Tourism's
Millennium Science Initiative through grant IC120009, awarded to The
Millennium Institute of Astrophysics (MAS). S.C.Y. was supported by
the Korea Astronomy and Space Science Institute under the R\&D
programme (Project No. 3348- 20160002) supervised by the Ministry of
Science, ICT and Future Planning and by the Monash Centre for
Astrophysics via the distinguished visitor programme.  E.~Y.~H. and
C.A. cknowledge the support provided by the National Science
Foundation under Grant No. AST-1613472 and by the Florida Space Grant
Consortium. F.F., J.M., G.M and A.R. acknowledge support from Conicyt
through the Fondecyt project No. 11130228. J.C.M., G.C-V., P.E., P.H.,
H.K., G.P. and F.O acknowledge support from Conicyt through the
Fondecyt projects No. 11170657, 3160747, 1171678, 3150460, 3140563,
1140352 and 11170953, respectively. G.M. and I.R. acknowledge support
from CONICYT-PCHA/MagisterNacional/2016-22162353 and 2016-22162464,
respectively.  G.G. is supported by the Deutsche
Forschungsgemeinschaft (DFG), Grant No. GR 1717/5.  S.G. acknowledges
support from Comit\'e Mixto ESO Chile project ORP 48/16. F.F., J.C.M.,
P.H., G.C-V. and P.E. acknowledge support from Conicyt through the
Programme of International Cooperation project DPI20140090. L.G. was
supported in part by the US National Science Foundation under Grant
AST-1311862. A.G.-Y. is supported by the EU via ERC grant No. 725161,
the Quantum Universe I-Core program, the ISF, the BSF Transformative
program and by a Kimmel award. M.F. is supported by the Royal
Society--Science Foundation Ireland University Research Fellowship
reference 15/RS-URF/3304. S.B. acknowledges funding from project RSCF
18-12-00522. Based on observations collected at the European
Organisation for Astronomical Research in the Southern Hemisphere
under ESO programmes 292.D-5042(A) and 094.D-0358(A). This work is
partly based on observations collected at the European Organisation
for Astronomical Research in the Southern Hemisphere, Chile as part of
PESSTO, (the Public ESO Spectroscopic Survey for Transient Objects
Survey) ESO program 188.D-3003, 191.D-0935, 197.D-1075. Partly based
on observations obtained with the GTC telescope.  This project used
data obtained with the Dark Energy Camera (DECam), which was
constructed by the Dark Energy Survey (DES) collaboration. Funding for
the DES Projects has been provided by the U.S. Department of Energy,
the U.S. National Science Foundation, the Ministry of Science and
Education of Spain, the Science and Technology Facilities Council of
the United Kingdom, the Higher Education Funding Council for England,
the National Center for Supercomputing Applications at the University
of Illinois at Urbana--Champaign, the Kavli Institute of Cosmological
Physics at the University of Chicago, Center for Cosmology and
Astro--Particle Physics at the Ohio State University, the Mitchell
Institute for Fundamental Physics and Astronomy at Texas A\&M
University, Financiadora de Estudos e Projetos, Funda\c{c}\~ao Carlos
Chagas Filho de Amparo, Financiadora de Estudos e Projetos,
Funda\c{c}\~ao Carlos Chagas Filho de Amparo \`a Pesquisa do Estado do
Rio de Janeiro, Conselho Nacional de Desenvolvimento Cient\'ifico e
Tecnol\'ogico and the Minist\'erio da Ci\^encia, Tecnologia e
Inova\c{c}\~ao, the Deutsche Forschungsgemeinschaft and the
Collaborating Institutions in the Dark Energy Survey. The
Collaborating Institutions are Argonne National Laboratory, the
University of California at Santa Cruz, the University of Cambridge,
Centro de Investigaciones Energ\'eticas, Medioambientales y
Tecnol\'ogicas–Madrid, the University of Chicago, University College
London, the DES--Brazil Consortium, the University of Edinburgh, the
Eidgen\"ossische Technische Hochschule (ETH) Z\"urich, Fermi National
Accelerator Laboratory, the University of Illinois at
Urbana--Champaign, the Institut de Ci\`encies de l'Espai (IEEC/CSIC),
the Institut de F\'isica d'Altes Energies, Lawrence Berkeley National
Laboratory, the Ludwig--Maximilians Universit\"at M\"unchen and the
associated Excellence Cluster Universe, the University of Michigan,
the National Optical Astronomy Observatory, the University of
Nottingham, the Ohio State University, the University of Pennsylvania,
the University of Portsmouth, SLAC National Accelerator Laboratory,
Stanford University, the University of Sussex, and Texas A\&M.
University.

\section*{Author contributions}
F.F. computed the supernova LCs based on DECam data and performed the
analysis presented in this work, including writing the analysis
software. T.J.M. computed the grid of supernova models used in this
work. F.F. and J.C.M. wrote the supernova discovery pipeline. J.P.A.,
F.B., A.C., Th.dJ., L.G., S.G-G., E.H., H.K., J.M., G.M., F.O., G.P.,
R.C.S. and A.K.V. helped with photometric and spectroscopic
observations under the HiTS program. T.J.M., S.B., G.G. and
S.C.Y. computed progenitor models. A.R. computed LC observations made
with the du Pont telescope. G.C-V., P.A.E., P.H., P.H., I.R. and
J.S.M. helped developing supernova detection algorithms, including
image processing, statistical methods and machine learning. R.C.S. and
E.V. coordinated the fast data access required to achieve the
real--time analysis and fast spectroscopic classifications. C.A.,
M.F., A.G-Y., E.K., L.LG., P.A.M., N.A.W. and D.R.Y. contributed to
PESSTO observations. A.dUP. contributed with spectroscopic
observations using the GTC telescope. All coauthors contributed with
comments. There are no competing interests in this work.

\newpage

\section*{Methods}

\paragraph{Classification spectra}
The following telegrams were sent to report the spectroscopic
observations and classifications performed for this project separated
by telescope: NTT
\cite{2014ATel.5957....1W,2014ATel.5970....1W,2015ATel.7144....1L,2015ATel.7154....1B},
SOAR
\cite{2015ATel.7291....1F,2015ATel.7246....1P,2015ATel.7164....1A,2015ATel.7335....1A}
and VLT \cite{2014ATel.6014....1A,2015ATel.7162....1A}. Two example
classification spectra are shown in Supplementary Figure~1 compared
to SN II spectra using SNID \cite{2007ApJ...666.1024B}.

\paragraph{Light curves}

The DECam data calibration included pre--processing, image difference,
candidate filtering and LC generation. For pre--processing the data we
used a modified version of the DECam community pipeline (DCP),
including electronic bias calibration, cross-talk correction,
saturation masking, bad pixel masking and interpolation, bias
calibration, linearity correction, flat-field gain calibration, fringe
pattern subtraction, bleed trail and edge bleed masking, and
interpolation. We removed cosmic rays using CRBlaster
\cite{2010PASP..122.1236M}. After pre--processing we used as templates
the first good quality images (i.e. with photometric conditions and
good seeing) of the fast cadence phase of observations when no other
templates were available. For fields which overlapped between 2014 and
2015 we used good quality images from 2014, which were deeper than in
2015. We aligned all the science images to the template using Lanczos
interpolation and we performed difference imaging using a variable
size pixel kernel as described in F16. Candidates were selected using
machine learning filters over stamps centered at pixels which reached
an integrated flux signal--to--noise ratio ($S/N$) above five.

To produce the LCs we first removed all flux differences which had a
science airmass larger than 1.7. Then, we improved the SN candidates
central position using all the available image differences. We
projected the empirical PSF to this new central position and performed
optimal photometry to estimate image difference fluxes and their
associated variances even when the object was not significantly
detected. Absolute calibrations were performed using photometric
nights as references and the zero points pre--computed for DECam at
different filters and CCDs. We have independently validated whether
these nights achieved photometric conditions using the PanSTARRS1
public catalogues \cite{2016arXiv161205243F}.

\paragraph{Spectral time series to light curves}

In order to ingest time series of synthetic spectra and produce
synthetic LCs for any redshift, attenuation and explosion time we
first assume a standard $\Lambda$--CDM cosmology. We redshift and
attenuate the spectral series with distance and assume a Cardelli law
with $R_V = 3.1$ for the dust attenuation. We integrate the resulting
spectra in the DECam bands and generate synthetics LCs which can be
interpolated into a given time array. In order to speed up the model
evaluation we pre--compute LCs for all DECam filters and all available
models in a logarithmically spaced time array, a linearly spaced
attenuation array and a logarithmically spaced redshift array. From
this different time, attenuation and redshift arrays we can
interpolate into given observational times assuming an explosion time,
an attenuation and a redshift.

\paragraph{Model Interpolation}

Apart from the previous interpolations we must be able to interpolate
quickly between models with different physical parameters. In order to
do this we first find the closest values in all the intrinsic physical
dimensions, e.g. mass, energy, mass loss rate $\dot M$ and wind
acceleration parameter $\beta$, and find all the models that have
combinations of these values, which we call $\vec{\theta}_{\rm
  close}$. The final LC will be a weighted combination of all these
models:
\begin{equation}
m(t, t_{\rm exp}, z, A_V, \vec{\theta}) = \sum_{\vec{\theta_i} \in
  {\vec{\theta}_{close}}} \hat{w}(\vec{\theta}, \vec{\theta_i}) ~m(t,
t_{\rm exp}, z, A_V, \vec{\theta_i}),
\end{equation}
where $m(t, t_{\rm exp}, z, A_V, \vec{\theta})$ is the magnitude of
the model at a given observation time $t$, explosion time $t_{\rm
  exp}$, redshift $z$, a given attenuation $A_V$ and a given vector of
model parameters $\vec{\theta}$; and the normalized weights
$\hat{w}(\vec{\theta}, \vec{\theta_i})$ are defined as:
\begin{equation}
\hat{w}(\vec{\theta}, \vec{\theta_i}) = \frac{w(\vec{\theta},
  \vec{\theta_i})}{\sum_{\vec{\theta_j} \in {\vec{\theta}_{close}}}
  w(\vec{\theta}, \vec{\theta_j})}
\end{equation}
where the weights are a function of a pair of parameter vectors
$\vec{\theta}$ and $\vec{\theta_i}$. In order to avoid having to
define a metric to compare values in the different dimensions of the
vector of physical parameters we define the weights to be inversely
proportional to the product of the differences in all the dimensions
of the vector of physical parameters $\vec{\theta}$, i.e.
\begin{equation}
 w(\vec{\theta}, \vec{\theta_i}) = \biggl( \prod_j |\theta^j - \theta_i^j| + \delta^j \biggr)^{-1},
\end{equation}
with $\vec{\delta}$ a vector with the same physical units as the
parameters, but much smaller than the typical separation in the grid
of models. This avoids the divergence of the weights when a given
coordinate matches the coordinates of known models.

For the attenuation, redshift and mass loss rate we use an internal
logarithmic representation. Since we also include models with $\dot M
= 0$, we assume them to correspond to a mass loss rate of $10^{-8}$
\msunyr.

The main advantages of the previous weighting scheme is that it does
not require defining a metric and it allows for possible missing
models in the grid. We show examples of interpolated models where we
vary the physical parameters smoothly, e.g. the mass loss rate
continuously between $10^{-8}$ and $10^{-2}$ \msunyr, in
Figure~\ref{fig:interp_mdot}.

\paragraph{Markov Chain Monte Carlo (MCMC) sampler}

Having the capability of quickly generating interpolated LCs for any
combination of explosion time, redshift, attenuation and vector of
physical parameters, we can approach the problem of inferring physical
parameters from a Bayesian perspective, i.e. computing the posterior
probability of the model parameters given the data and assuming prior
distributions.

To do this we sample the posterior of the joint distribution of
parameters using a Markov Chain Monte Carlo (MCMC) sampler which uses
an affine invariant approach \cite{2010CAMCS...5...65G}. This method uses
parallel Markov chains that sample the posterior distribution moving
randomly in directions parallel to the relative positions of the
samplers, following acceptance rules that satisfy the condition of
detailed balance for reversible Markov chains. This is implemented in
python via \emph{emcee} \cite{2013PASP..125..306F}.

We run the MCMC sampler using pre--defined priors which are relatively
flat distributions for most variables in linear (mass, energy,
$\beta$) or logarithmic (redshift, $A_V$, $\dot M$) scale (see
Supplementary Table~3). For the explosion time we require a first
guess, for which we run an interactive fitting routine for all the SN
LCs using tools found in our public
\href{https://github.com/fforster/surveysim}{repository}, and use a
Gaussian prior around this value with a standard deviation of 4
days. We also allow for a variable scale parameter to allow for errors
in absolute calibrations, for which we use a Gaussian prior centered
at 1.0 and with a standard deviation of 0.01 (1\% errors). We use 400
parallel samplers (or \emph{walkers}) and 900 steps per sampler, with
a burn--in period of 450 steps in all cases. These numbers were set
via trial and error, trying to reach the detailed balance condition
while reducing the number of multiple disjoint solutions in the
sampler. The Markov chains reached acceptance ratios between 0.2 and
0.5 in more than 95\% of the cases.  An example joint distribution can
be visualized in a corner plot as shown in Supplementary
Figure~2 and example light curves drawn from the
posterior distribution are shown for the HiTS SN II sample (Figure
\ref{fig:LCs}) and SNe II in the literature with well sampled optical
early light curves (Figure \ref{fig:MCMCother}).

In order to extract posterior probabilities for a given parameter we
can marginalize the joint distribution integrating over certain
dimensions, e.g. explosion time, redshift when not available,
attenuation $A_V$, mass, and energy. This is important in order to
derive conclusions about the true distributions of $\dot M$ and
$\beta$.

\paragraph{Photometric classifier examples} In Supplementary Figure~3 we show two observed light curves
compared to posterior sampled light curves assuming the SN II or
scaled SN Ia models discussed in the text.

\paragraph{Selection effects}

Given the different filters applied to the data it is reasonable to
ask whether they play any role in the observed excess of SNe II with
$\dot M > 10^{-4}$ \msunyr. Two cases need to be considered:
contamination of other SN types at large $\dot M$ values, or a lower
detection/classification efficiency of SNe II with low $\dot M$. The
former seems unlikely, since no other SN types show the very fast rise
times observed in our sample (contamination from other SN types at low
$\dot M$ values seems more likely, but this would increase the number
of low $\dot M$ SN II candidates, in the opposite direction of what we
observe), but the latter cannot be discarded a priori. Low
detection/classification efficiencies at low $\dot M$ could be due to
four possibilities: 1) lower detection efficiency of low $\dot M$ SNe
II, i.e. that relatively fewer SNe II with low $\dot M$ reach the
necessary $S/N$ for detection; 2) lower classification efficiency,
i.e. that low $\dot M$ SNe II are more often miss--classified as Type
I SNe by our classifier than large $\dot M$ SNe II; 3) lower selection
efficiency when removing SNe with a poorly sampled rise, i.e. that
either removing SNe with gaps in the data or with a non--negligible
probability of having been seen three or more days after their first
light favors large $\dot M$ SNe; or 4) that the inferrence process is
biased towards large $\dot M$ values.

We have investigated all the previous possibilities via simulations
assuming a uniform logarithmic distribution of $\dot M$ values between
$10^{-8}$ and $10^{-2}$ \msunyr\ and testing whether there is an
excess in the recovery fractions of the samples below and above the
median value of $10^{-5}$ \msunyr, which we call the low and large
$\dot M$ samples, respectively. First, for case 1) we simulate
$150,000$ synthetic SNe II with a uniform distribution of $\log_{10}
\dot M$, $\beta$, mass, energy, explosion time and an exponential
distribution of attenuations, taking into account cosmology, the star
formation history, the efficiency at converting stars into SNe II and
the actual cadence and depth of the survey (see F16 for more
details). We found that the number of SNe with at least two detections
(with $S/N \ge 5$) and which have at least one observation (with $S/N
\ge 2$) within the first three days after emergence should be
distributed in a proportion of 28 to 72\% between the low and large
$\dot M$ samples, i.e. there is a significant bias towards large $\dot
M$ values. However, using a binomial distribution with the expected
fractions of low and large $\dot M$ SNe II and the total number of SNe
in the sample we infer that the probability of having two or less SNe
in the low $\dot M$ sample is only 1\%. Moreover, the probability of
having two or less SNe below $10^{-4}$ \msunyr\ is only $2 \times
10^{-4}$, i.e. we can discard that the relative absence of low $\dot
M$ SNe II is due to a statistical fluctuation. With these simulations
we can also compute the detection efficiency as a function of
$\log_{10} \dot M$ (Supplementary Figure~4), which we will use to
correct the sample distribution of $\log_{10} \dot M$ values.

In order to test case 2) we simulated 300 SNe whose parameters are
drawn from the inferred distribution of parameters in our sample,
except for the mass loss rate which is drawn from a uniform
logarithmic distribution, and run MCMC to get median log--likelihoods
and the BIC classifier on them. We have found that more than 95\% of
the SNe in the low $\dot M$ sample are correctly classified as SNe II,
therefore this effect cannot explain the relative absence of low $\dot
M$ SN II candidates. For case 3) we expect that the presence of gaps
in the data during the initial rise should not be related to the value
of $\dot M$, but perhaps SNe with low $\dot M$ values have a more
uncertain time of emergence (due to their shallower rise) which would
lead to larger reported probabilities of having missed the initial
three days after first light. We tested this case by measuring the
relative uncertainties (percentile 50 minus percentile 10) in the
inferred time of first light in the low and large $\dot M$ samples,
finding that in the large $\dot M$ case uncertainties tend to be 50\%
smaller. Therefore, there could be some low $\dot M$ SNe occuring at
the beginning of the survey which would be preferentially
removed. Consider the case when a low $\dot M$ SN had its first light
at the time of first observation, but its uncertainties allowed for a
first light more than three days before explosion (for comparison, the
median and maximum difference between the percentiles 50 and 5 of the
inferred explosion times are 1.8 and 2.9 days, respectively). A
similarly observed large $\dot M$ SN having its first light at the
same time would not have been removed because its uncertainties were
50\% smaller, giving this family of explosions up to one additional
effective survey day, or a 10\% relative excess considering
Supplementary Figure~5. This cannot explain the large observed excess
of large $\dot M$ SN II candidates either. Finally, for case 4) we
performed similar simulations with a flat distribution of log mass
loss rates and found that if anything there is a bias towards small
$\dot M$ values (see discussion in following paragraphs). Thus, from
the discussion above we conclude that the excess of large $\dot M$ SN
II candidates is not due to a selection effect and that it is
significant.

\paragraph{Parameter inference tests}

In Supplementary Figure~6 we show a kernel density estimation of
the sum of posterior distributions marginalized over $\dot M$ and
$\beta$ for the 26 SNe in the sample. In order to combine
distributions with very different variances we use Silverman's rule
to estimate a kernel width \cite{1986desd.book.....S}:
\begin{align}
	h = 0.9 ~ \min(\sqrt{{\rm var}(x)}, ~{\rm IQR}(x) / 1.349) ~
  N^{1/5},
\end{align}
where $h$ is the kernel width, $x$ is the random variable whose
distribution we want to estimate, var is the variance, IQR is the
inter--quartile range and $N$ is the number of sampled values of
$x$. In our case we use the sampled posterior as our random variable,
marginalizing over all other variables first. We correct these
distributions by the detection efficiencies derived previously,
confirming that $\dot M < 10^{-4}$ \msunyr\ is not favored by the
data. Although we find some significantly large values of $\beta$ for
individual SNe, we do not find a strong preference for large $\beta$
values in the sample.

In order to test whether the preference for large $\dot M$ could be
due to some bias in the posterior sampling we simulate 300 SNe from a
uniform distribution in $\dot M$ in logarithmic scale, a uniform
distribution in $\beta$, and a uniform distribution in progenitor
masses. We use the inferred explosion times from the observed SN
sample to mimic the same time coverage; and the inferred energy,
redshift and attenuation to mimic the observed apparent magnitudes. We
then run the same posterior sampling algorithm and test whether the
sum of the posterior distributions resembles the input
distribution. For $\dot M$ we see a relatively flat recovered
distribution with a median value at $10^{-5.4}$ \msunyr,
i.e. below $10^{-6}$ \msunyr\ or only a slight bias towards small mass
loss rates. For $\beta$ we also see a relatively flat distribution
with a median $\beta$ of 2.8, below 3.0 or a slight bias towards small
values as well. For comparison, the posterior distribution of $\dot M$
in the observed sample has a median value of $10^{-2.8}$ \msunyr\ and
the posterior distribution of $\beta$ in the observed sample has a
median value of 3.5.

Then, in order to test whether there is a bias in $\beta$ or the
progenitor masses at the range of high $\dot M$ suggested by our
observations, we simulated another 300 SNe where we instead use the
inferred distribution of $\dot M$, but uniform distributions of
$\beta$ and progenitor mass. Again, we do not find a bias in $\beta$,
with a relatively flat distribution and a median at 3.0, exactly in
the middle of the distribution. However, we detect a large bias in the
progenitor mass distribution towards small values, with a median
inferred progenitor mass in both the simulated and observed sample of
12.8 $M_\odot$ (instead of 14.0 $M_\odot$). This means that we are not
able to derive meaningful conclusions about the distribution of
progenitor masses in the sample. For the case of energy, where we
expect a strong degeneracy with redshift, we also simulated a random
sample of 300 SNe using inferred values for all the physical
quantities, except for the energy which is drawn from a uniform
distribution between 0.5 and 2. The median of the inferred
distribution is at 1.30, slightly above 1.25, which means that there
is only a slight bias towards larger energies. However, the standard
deviation of the differences between the medians of the inferred
energy distribution and the simulated energies is 0.44, almost the
same as the value one would get assuming flat energy posteriors
(0.43). Thus, the energy is poorly constrained by our observations,
but we detect no significant bias. Note that although energy is poorly
constrained, it is important to marginalize over its possible values
to learn about the distributions of the other variables.

Using simulated light curves we can also test how well we recover each
parameter. For this we measure the root mean square (r.m.s.)
difference between simulated values and the median of the posterior
distribution for each SNe. We obtain an r.m.s. of 1.2 for $\log_{10}
\dot M$, 1.4 for $\beta$, 1.2 $M_\odot$ for the progenitor mass, 0.03
for the redshift, 0.44 for the energy (see previous paragraph) and
0.14 for the attenuation. Interestingly, we found that the three SNe
with host galaxy redshifts had values consistent with those inferred
from the light curves alone, i.e. allowing for the redshift to vary
during Bayesian inferrence. The host galaxy redshifts for SNHiTS15C,
SNHiTS15aq and SNHiTS15aw are 0.08, 0.11, 0.07, respectively, while
the 5, 50 and 95 redshift percentiles inferred from the light curves
alone are 0.08, 0.10, 0.12; 0.10, 0.11, 0.13; and 0.06, 0.07, 0.08;
respectively.

We also test whether some of the inferred variables are
correlated. For this we build a correlation matrix using the
\emph{median} values of the posterior distributions. We found that the
only two variables that are significantly correlated are $A_V$ and
$\log_{10} \dot M$, with a correlation coefficient of -0.79, which
points to a degeneracy in the normalization of the light curves. If we
remove the two low mass loss rate SNe in the sample, SNHiTS15G and
SNHiTS15Q, this correlation weakens to -0.5 and the stronger
correlations become redshift with $\log_{10} \dot M$ (0.68) and
redshift with $\beta$ (-0.66), possibly pointing to a degeneracy in
the shape of the LCs (see Figure~\ref{fig:interp_mdot}). No other
off--diagonal correlation coefficients are larger than 0.6 in absolute
value. This suggests that having independent redshift determinations
for the SNe in the sample would significantly improve the quality of
our results, as exemplified by SNHiTS15aw.

In order to test for biases related to the observational strategy we
study the distribution of explosion times for the HiTS14A and HiTS15A
campaigns (see F16), which we show in Supplementary Figure~5. For the
2014A campaign the posterior of explosion times span a shorter time
than the survey itself, which suggests some bias towards explosions
happening at the beginning of the survey. For the 2015A campaign we
detect a possible bias for detections during the initial high cadence
phase of observations as well. The difference between the maximum and
minimum median explosion time is 9.5 days, and the median of these
values differs by 1.3 days from the middle point between the minimum
and maximum values.  We observe a similar behaviour when estimating
detection efficiencies as a function of explosion time.

We also show the distribution of inferred redshifts and attenuations
in Supplementary Figure~7. We can see that the survey efficiency starts
decreasing at redshift 0.3, and that we may be able to detect SNe up
to redshift 0.4--0.5, which highlights the potential for similar
surveys to be used for high redshift SNe II studies. The apparent
bimodality is not significant: we run a Hartigans' Dip test of
Unimodality \cite{HH85} using the inferred median redshift values and
we could not reject unimodality with a $p$--value of 0.84.  The
distribution of attenuations appears to follow an exponential
distribution with a characteristic scale of 0.07. It is worth noticing
that in our simulations most SNe above redshift of 0.3 have $\dot M >
10^{-3}$ \msunyr, which explains the previously described selection
effects.

Finally, we show the distribution of favored $\dot M$ values as a
function of redshift in Supplementary Figure~8, where a lack of low $\dot
M$ models at high redshifts is observed.  In the same figure we show
the predicted distribution of $\log_{10} \dot M$ and redshift as
linearly spaced density contours, assuming a flat distribution of
$\log_{10} \dot M$. These simulations predict more SNe II than
observed at both very large mass loss rates (only two SNe with $\dot M
> 10^{-2.45}$ \msunyr) and at low mass loss rates (only two SNe with
$\dot M < 10^{-4}$ \msunyr), with probabilities of 0.07 and 0.0002,
respectively, i.e. they are unlikely to be due to a statistical
fluctuation.

\section*{Data availability}


The data that support the plots within this paper and other findings
of this study are available from the corresponding author upon reasonable request.

\section*{Bibliography}

\let\oldthebibliography=\thebibliography
\let\oldendthebibliography=\endthebibliography
\renewenvironment{thebibliography}[1]{%
  \oldthebibliography{#1}%
  \setcounter{enumiv}{56}%
}{\oldendthebibliography}

\beginmethods

\begin{table*}[ht!]
  \footnotesize
  \begin{centering}
    \begin{tabular}{l|ccc|ccc|ccc}
      \hline
      SN name & \multicolumn{3}{|c|}{$z$} & \multicolumn{3}{|c|}{$A_V$} & \multicolumn{3}{|c}{$t_{\rm exp}\ [MJD]$} \\
      \hline  
\vspace{-.2cm}      SNHiTS14B & 0.18 & 0.22 & 0.27 & 0.00 & 0.03 & 0.27 & 56713.57 & 56715.32 & 56716.40 \\
\vspace{-.2cm}      SNHiTS14C & 0.08 & 0.08 & 0.08 & 0.01 & 0.09 & 0.25 & 56712.12 & 56713.78 & 56715.13 \\
\vspace{-.2cm}      SNHiTS14N & 0.18 & 0.28 & 0.39 & 0.00 & 0.04 & 0.46 & 56712.42 & 56715.02 & 56716.50 \\
\vspace{-.2cm}      SNHiTS14Q & 0.20 & 0.25 & 0.31 & 0.00 & 0.03 & 0.27 & 56713.37 & 56715.44 & 56716.64 \\
\vspace{-.2cm}      SNHiTS14T & 0.16 & 0.20 & 0.23 & 0.00 & 0.06 & 0.25 & 56713.32 & 56714.32 & 56715.16 \\
\vspace{-.2cm}      SNHiTS15A & 0.21 & 0.25 & 0.29 & 0.00 & 0.03 & 0.26 & 57064.48 & 57066.18 & 57067.38 \\
\vspace{-.2cm}      SNHiTS15D & 0.13 & 0.16 & 0.18 & 0.00 & 0.02 & 0.17 & 57065.45 & 57066.98 & 57067.56 \\
\vspace{-.2cm}      SNHiTS15F & 0.20 & 0.24 & 0.28 & 0.00 & 0.02 & 0.20 & 57063.59 & 57064.97 & 57065.99 \\
\vspace{-.2cm}      SNHiTS15G & 0.10 & 0.14 & 0.17 & 0.00 & 0.04 & 0.26 & 57067.11 & 57068.07 & 57069.12 \\
\vspace{-.2cm}      SNHiTS15K & 0.18 & 0.22 & 0.29 & 0.00 & 0.03 & 0.32 & 57063.37 & 57065.48 & 57067.49 \\
\vspace{-.2cm}      SNHiTS15M & 0.28 & 0.37 & 0.45 & 0.00 & 0.03 & 0.30 & 57063.74 & 57065.93 & 57068.62 \\
\vspace{-.2cm}      SNHiTS15P & 0.22 & 0.24 & 0.26 & 0.00 & 0.02 & 0.11 & 57070.34 & 57070.77 & 57071.08 \\
\vspace{-.2cm}      SNHiTS15Q & 0.04 & 0.05 & 0.06 & 0.13 & 0.67 & 0.80 & 57069.87 & 57070.41 & 57071.18 \\
\vspace{-.2cm}      SNHiTS15X & 0.23 & 0.31 & 0.39 & 0.00 & 0.03 & 0.26 & 57066.01 & 57068.43 & 57070.32 \\
\vspace{-.2cm}      SNHiTS15ag & 0.29 & 0.42 & 0.52 & 0.00 & 0.04 & 0.44 & 57063.67 & 57065.79 & 57068.13 \\
\vspace{-.2cm}      SNHiTS15ah & 0.32 & 0.41 & 0.49 & 0.00 & 0.03 & 0.27 & 57066.58 & 57067.27 & 57070.52 \\
\vspace{-.2cm}      SNHiTS15ai & 0.24 & 0.34 & 0.43 & 0.00 & 0.03 & 0.39 & 57065.97 & 57068.09 & 57070.15 \\
\vspace{-.2cm}      SNHiTS15ak & 0.21 & 0.31 & 0.39 & 0.00 & 0.03 & 0.32 & 57067.00 & 57069.09 & 57070.83 \\
\vspace{-.2cm}      SNHiTS15aq & 0.10 & 0.11 & 0.13 & 0.00 & 0.02 & 0.15 & 57070.58 & 57072.75 & 57072.96 \\
\vspace{-.2cm}      SNHiTS15as & 0.19 & 0.26 & 0.33 & 0.00 & 0.03 & 0.38 & 57063.60 & 57065.50 & 57066.98 \\
\vspace{-.2cm}      SNHiTS15aw & 0.07 & 0.07 & 0.07 & 0.17 & 0.25 & 0.31 & 57074.40 & 57074.47 & 57074.89 \\
\vspace{-.2cm}      SNHiTS15ay & 0.25 & 0.28 & 0.31 & 0.00 & 0.02 & 0.17 & 57068.72 & 57069.38 & 57069.72 \\
\vspace{-.2cm}      SNHiTS15az & 0.23 & 0.28 & 0.38 & 0.00 & 0.02 & 0.21 & 57064.66 & 57067.64 & 57070.75 \\
\vspace{-.2cm}      SNHiTS15bc & 0.33 & 0.43 & 0.55 & 0.00 & 0.03 & 0.22 & 57068.28 & 57071.15 & 57072.59 \\
\vspace{-.2cm}      SNHiTS15bm & 0.15 & 0.20 & 0.27 & 0.00 & 0.07 & 0.62 & 57071.44 & 57073.21 & 57074.49 \\
\vspace{-.2cm}      SNHiTS15ch & 0.16 & 0.21 & 0.26 & 0.00 & 0.03 & 0.28 & 57069.92 & 57072.19 & 57073.73 \vspace{.2cm} \\
\hline
\vspace{-.2cm}      SN2006bp & 0.004 & 0.005 & 0.007 & 0.10 & 0.68 & 1.24 & 56387.40 & 56390.17 & 56392.12 \\
\vspace{-.2cm}      PS1-13arp & 0.17 & 0.17 & 0.17 & 0.00 & 0.09 & 0.34 & 56387.40 & 56390.17 & 56392.12 \\
\vspace{-.2cm}      SN2013fs & 0.01 & 0.01 & 0.01 & 0.11 & 0.31 & 0.32 & 56566.81 & 56567.33 & 56567.35\\
\vspace{-.2cm}      KSN2011a & 0.05 & 0.05 & 0.05 & 0.00 & 0.02 & 0.10 & 55762.86 & 55763.28 & 55763.28 \\
\vspace{-.2cm}      KSN2011d & 0.09 & 0.09 & 0.09 & 0.01 & 0.05 & 0.10 & 55868.76 & 55869.19 & 55869.34 
    \end{tabular}
    \vspace{.5cm}
    \caption{Percentiles 5, 50 and 95 of the posterior distribution of
      physical parameters marginalized over the redshift $z$,
      attenuation $A_V$ and explosion time $t_{\rm exp}$ for the
      sample of 26 HiTS SN II candidates and 5 SNe II from the
      literature. For SN2006bp, the most nearby SN in the sample ($z =
      0.0035$), we allow for a variable redshift to account for
      possible host proper motions.}
  \end{centering}
  \label{suptab:allsne1}
\end{table*}

\begin{table*}[ht!]
  \footnotesize
  \begin{centering}
    \begin{tabular}{l|ccc|ccc|ccc|ccc}
      \hline
      SN name & \multicolumn{3}{|c|}{$\log_{10}\dot M$ [\msunyr]} & \multicolumn{3}{|c|}{$\beta$} & \multicolumn{3}{|c|}{mass [$M_\odot$]} & \multicolumn{3}{|c}{energy [B]} \\
      \hline 
\vspace{-.2cm}      SNHiTS14B & -3.09 & -2.70 & -2.52 & 1.13 & 3.58 & 4.84 & 12.05 & 13.05 & 14.78 & 0.62 & 1.06 & 1.62 \\
\vspace{-.2cm}      SNHiTS14C & -3.44 & -2.85 & -2.68 & 1.52 & 3.38 & 4.88 & 12.06 & 12.75 & 14.74 & 0.53 & 0.72 & 1.04 \\
\vspace{-.2cm}      SNHiTS14N & -4.03 & -3.13 & -2.53 & 1.44 & 3.32 & 4.83 & 12.11 & 13.28 & 15.54 & 0.57 & 1.14 & 1.90 \\
\vspace{-.2cm}      SNHiTS14Q & -3.08 & -2.66 & -2.17 & 1.37 & 3.12 & 4.73 & 12.04 & 12.55 & 14.37 & 0.65 & 1.35 & 1.90 \\
\vspace{-.2cm}      SNHiTS14T & -3.11 & -2.85 & -2.56 & 1.51 & 3.28 & 4.72 & 12.06 & 12.68 & 14.30 & 1.05 & 1.61 & 1.96 \\
\vspace{-.2cm}      SNHiTS15A & -3.56 & -3.43 & -2.98 & 2.38 & 3.72 & 4.90 & 12.11 & 13.75 & 15.71 & 1.38 & 1.80 & 1.99 \\
\vspace{-.2cm}      SNHiTS15D & -3.59 & -3.43 & -2.95 & 2.11 & 3.81 & 4.89 & 12.11 & 12.71 & 14.89 & 0.82 & 1.38 & 1.96 \\
\vspace{-.2cm}      SNHiTS15F & -3.06 & -2.92 & -2.75 & 2.43 & 3.81 & 4.90 & 12.10 & 13.05 & 14.61 & 0.83 & 1.74 & 1.98 \\
\vspace{-.2cm}      SNHiTS15G & -7.04 & -5.86 & -4.48 & 1.64 & 2.62 & 4.49 & 13.53 & 14.92 & 15.92 & 0.64 & 1.32 & 1.92 \\
\vspace{-.2cm}      SNHiTS15K & -3.27 & -2.83 & -2.58 & 1.56 & 3.50 & 4.88 & 12.23 & 13.80 & 15.69 & 0.54 & 0.97 & 1.85 \\
\vspace{-.2cm}      SNHiTS15M & -2.90 & -2.33 & -2.08 & 1.29 & 2.83 & 4.65 & 12.11 & 13.14 & 15.44 & 0.79 & 1.44 & 1.93 \\
\vspace{-.2cm}      SNHiTS15P & -3.00 & -2.89 & -2.70 & 3.25 & 3.76 & 4.76 & 12.01 & 12.09 & 12.44 & 1.43 & 1.72 & 1.97 \\
\vspace{-.2cm}      SNHiTS15Q & -7.98 & -7.70 & -7.16 & 2.07 & 3.37 & 4.68 & 12.01 & 12.09 & 12.38 & 0.55 & 0.92 & 1.35 \\
\vspace{-.2cm}      SNHiTS15X & -3.00 & -2.55 & -2.11 & 1.26 & 2.90 & 4.71 & 12.13 & 13.28 & 15.45 & 0.74 & 1.40 & 1.92 \\
\vspace{-.2cm}      SNHiTS15ag & -2.85 & -2.41 & -2.14 & 1.24 & 2.85 & 4.72 & 12.15 & 13.53 & 15.61 & 0.74 & 1.41 & 1.93 \\
\vspace{-.2cm}      SNHiTS15ah & -3.01 & -2.49 & -2.21 & 1.20 & 2.56 & 4.51 & 12.05 & 12.65 & 14.58 & 1.04 & 1.62 & 1.97 \\
\vspace{-.2cm}      SNHiTS15ai & -3.45 & -2.76 & -2.42 & 1.25 & 2.97 & 4.85 & 12.19 & 13.82 & 15.76 & 0.64 & 1.23 & 1.87 \\
\vspace{-.2cm}      SNHiTS15ak & -3.59 & -2.82 & -2.45 & 1.27 & 3.15 & 4.84 & 12.22 & 13.91 & 15.75 & 0.60 & 1.22 & 1.88 \\
\vspace{-.2cm}      SNHiTS15aq & -3.52 & -3.40 & -2.67 & 3.45 & 3.80 & 4.98 & 12.01 & 12.18 & 13.28 & 0.63 & 1.31 & 1.83 \\
\vspace{-.2cm}      SNHiTS15as & -3.55 & -3.17 & -2.80 & 1.56 & 3.65 & 4.89 & 12.11 & 13.14 & 15.38 & 0.67 & 1.40 & 1.93 \\
\vspace{-.2cm}      SNHiTS15aw & -3.57 & -3.51 & -3.42 & 3.69 & 3.75 & 3.82 & 12.04 & 12.56 & 12.85 & 0.67 & 0.78 & 0.87 \\
\vspace{-.2cm}      SNHiTS15ay & -3.04 & -2.99 & -2.91 & 2.35 & 3.23 & 4.40 & 12.01 & 12.16 & 12.66 & 1.60 & 1.90 & 1.99 \\
\vspace{-.2cm}      SNHiTS15az & -3.55 & -2.69 & -2.09 & 1.54 & 3.63 & 4.91 & 12.09 & 13.24 & 15.56 & 0.52 & 0.92 & 1.88 \\
\vspace{-.2cm}      SNHiTS15bc & -3.29 & -2.81 & -2.25 & 1.36 & 3.10 & 4.76 & 12.05 & 12.65 & 14.69 & 0.82 & 1.63 & 1.97 \\
\vspace{-.2cm}      SNHiTS15bm & -4.39 & -3.22 & -2.54 & 1.24 & 2.85 & 4.76 & 12.31 & 14.11 & 15.77 & 0.57 & 1.10 & 1.79 \\
\vspace{-.2cm}      SNHiTS15ch & -3.78 & -3.09 & -2.43 & 1.35 & 3.40 & 4.86 & 12.31 & 14.10 & 15.79 & 0.58 & 1.12 & 1.86 \vspace{.2cm} \\
\hline
\vspace{-.2cm}      SN2006bp & -3.13 & -2.68 & -2.58 & 3.74 & 4.69 & 5.00 & 12.80 & 13.28 & 13.53 & 1.58 & 1.63 & 1.71 \\
\vspace{-.2cm}      PS1-13arp & -3.35 & -2.93 & -2.60 & 1.30 & 3.42 & 4.81 & 12.06 & 12.84 & 14.79 & 0.54 & 0.70 & 0.91 \\
\vspace{-.2cm}      SN2013fs & -2.71 & -2.62 & -2.61 & 3.75 & 3.75 & 4.48 & 12.00 & 12.00 & 12.16 & 1.02 & 1.33 & 1.35 \\
\vspace{-.2cm}      KSN2011a & -2.73 & -2.60 & -2.52 & 1.08 & 2.49 & 3.00 & 12.00 & 12.06 & 12.19 & 0.60 & 0.65 & 0.87 \\
\vspace{-.2cm}      KSN2011d & -2.77 & -2.72 & -2.58 & 4.08 & 4.40 & 4.66 & 15.07 & 15.34 & 15.64 & 1.22 & 1.28 & 1.34
    \end{tabular}
    \vspace{.5cm}
    \caption{Percentiles 5, 50 and 95 of the posterior distribution of
      physical parameters marginalized over the mass loss rate $\dot
      M$, wind acceleration parameter $\beta$, mass and energy for the
      sample of 26 HiTS SN II candidates and 5 SNe II from the
      literature. }
  \end{centering}
  \label{suptab:allsne2}
\end{table*}

\begin{table}
  \begin{centering}
    \begin{tabular}{cc}
      Variable $\sim$ prior distribution & units\\
      \hline
      $t_{\rm exp}$ $\sim$ N($t_{\rm exp}^{\rm guess}$, 4) & days \\
      $\ln z$ $\sim$ N($\ln 0.18$, 2), $z \in (10^{-3}, 1)$ & \\
      $\ln A_{\rm V}$ $\sim$ N($\ln 0.05$, 2), $A_{\rm V} \in (10^{-4}, 10)$ & mag. \\
      mass $\sim$ N(14, 3), mass $\in (12, 16)$ & $M_\odot$ \\
      energy $\sim$ N(1, 1), energy $\in (0.5, 2)$ & B \\
      $\log_{10} \dot M$ $\sim$ U(-8, 2), $\log_{10} \dot M \in (-8, -2)$ & \msunyr \\
      $\beta$ $\sim$ N(3, 2), $\beta \in (0.5, 2)$ & \\
    \end{tabular}
    \caption{Prior distributions for different variables used in the
      analysis and their units. The distributions are Gaussian,
      $N(\mu, \sigma)$ is a Gaussian distribution with mean $\mu$ and
      standard deviation $\sigma$; or uniform, $U(a, b)$ is a uniform
      distribution between $a$ and $b$. The prior probabilities are
      zero outside the intervals indicated. $t_{\rm exp}^{\rm guess}$
      is a first guess explosion time.}
    \label{suptab:priors}
  \end{centering}
\end{table}

\begin{figure}[ht!]
  \begin{centering}
    \hbox{
      \includegraphics[width=0.5\hsize, bb=0 0 686 427, angle=0]{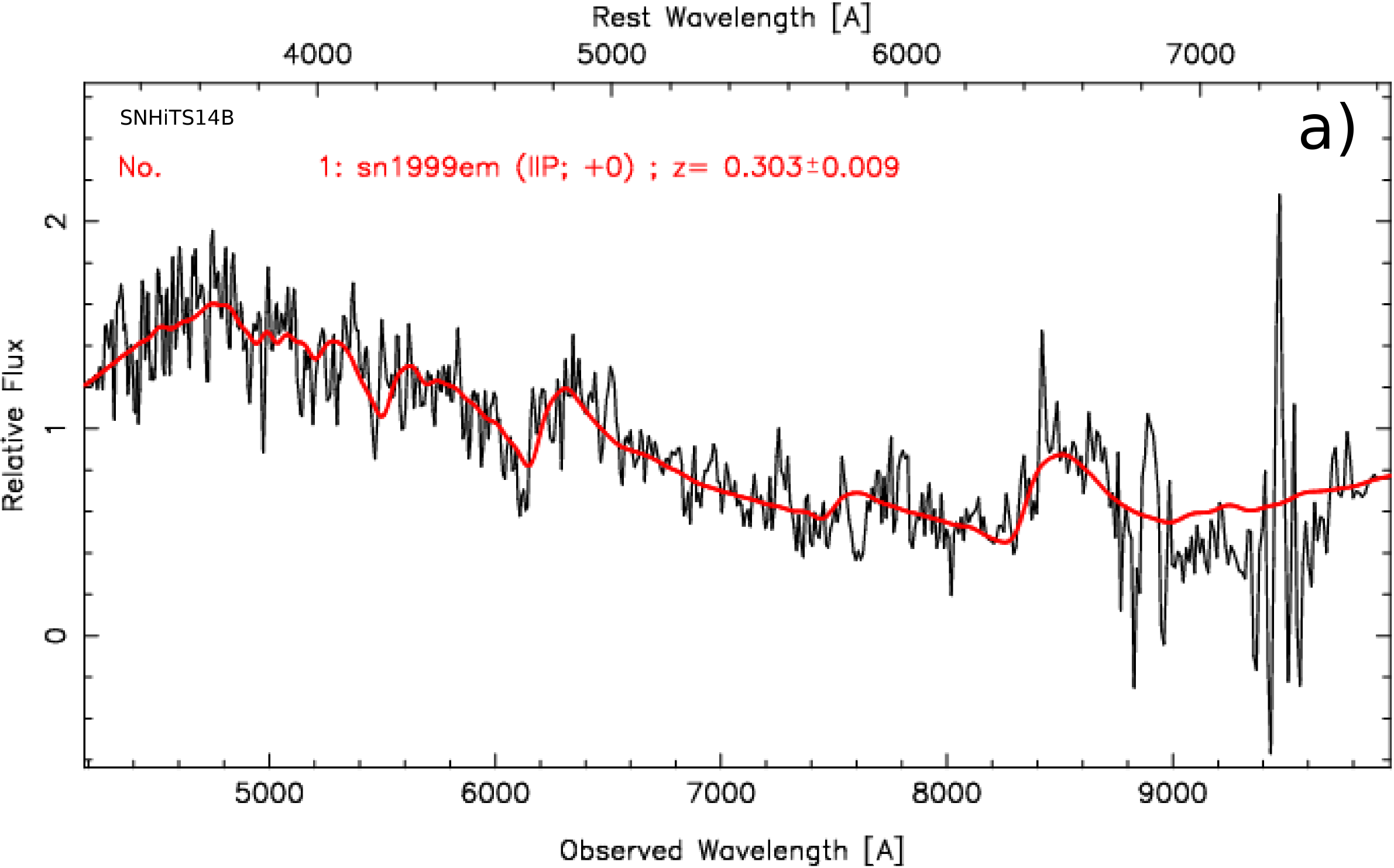}
      \includegraphics[width=0.5\hsize, bb=0 0 686 427, angle=0]{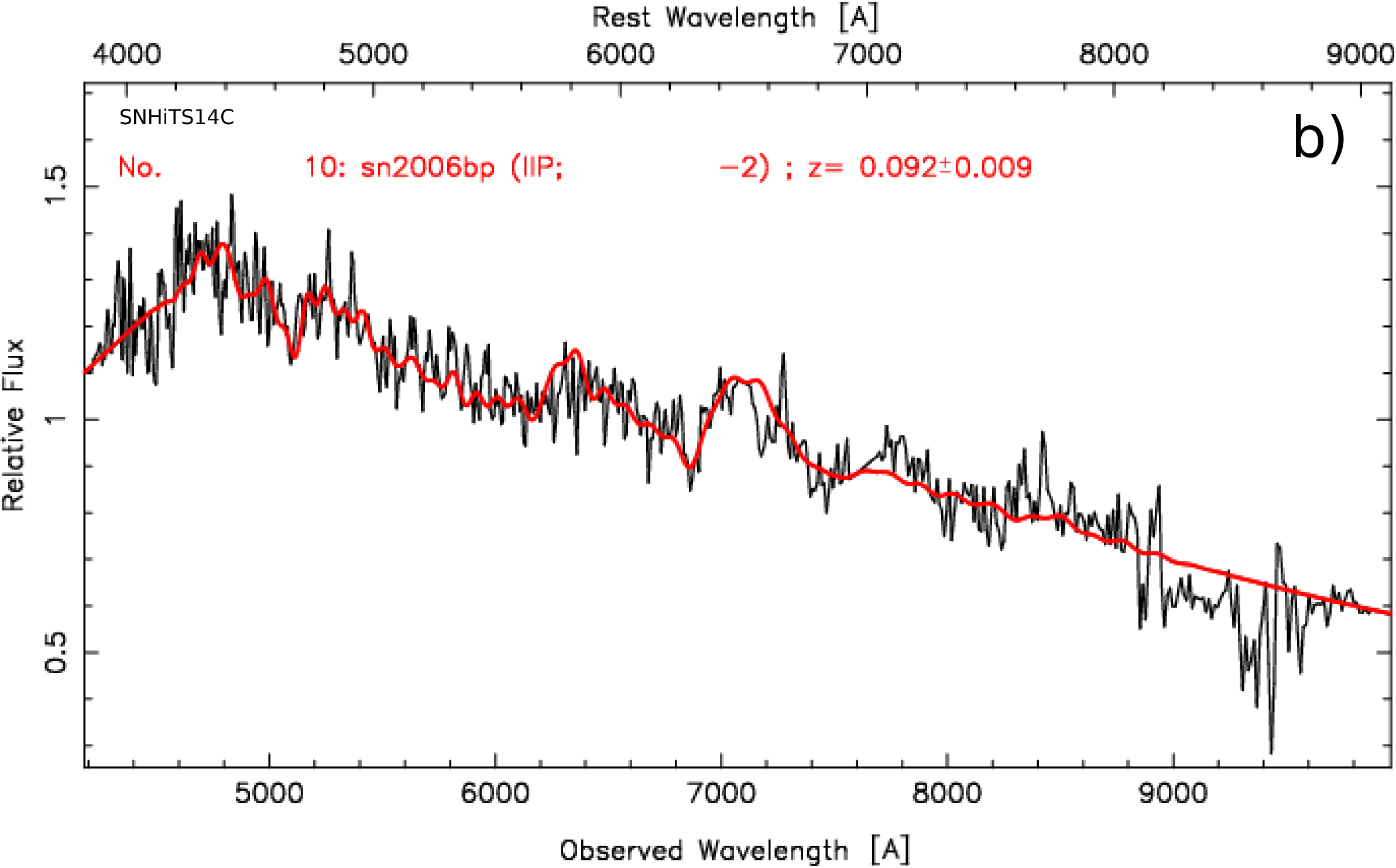}
    }
    \caption{Classification spectra of SNHiTS14B (a) and SNHITS14C (b)
      compared to SNID \cite{2007ApJ...666.1024B} templates.}
    \label{supfig:spectra}
  \end{centering}
\end{figure}

\begin{figure}[ht!]
\includegraphics[width=0.9\hsize, bb=0 0 1200 1200]{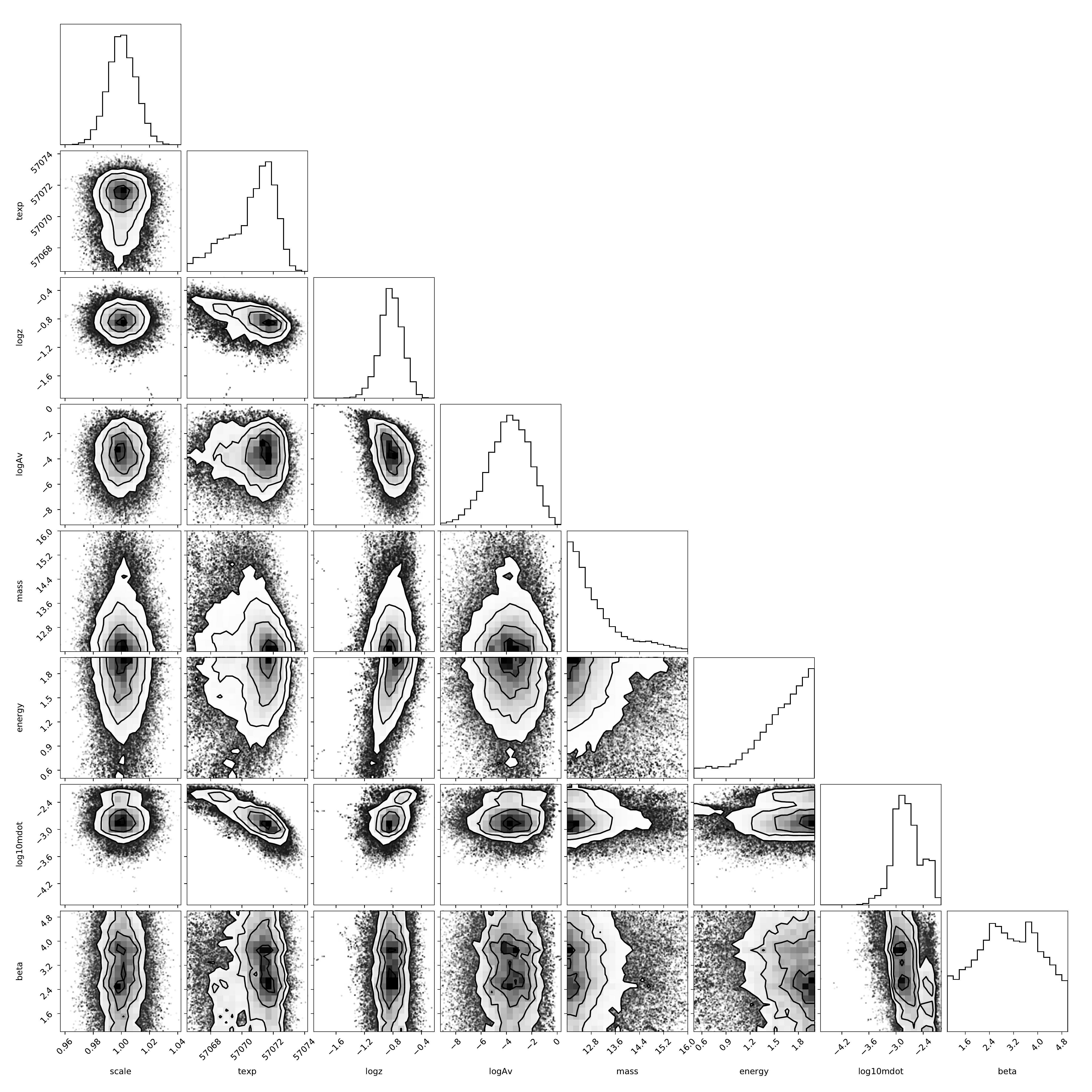}
\caption{Corner plot visualization of the posterior joint probability
  distribution of the parameters given the LC of Type II SN candidate
  SNHiTS15bc. The parameters used for this problem were a scale factor
  (forced to be close to one via a prior), the explosion time, the
  redshift $z$, the attenuation $A_V$, the main sequence mass, the
  energy of the explosion, the mass loss rate $\dot M$ and the wind
  acceleration parameter $\beta$. Note that $\dot M < 10^{-4}$ \msunyr\ 
  are not favored for any combination of the other parameters.}
\label{supfig:corner}
\end{figure}

\begin{figure*}[ht!]
  \vbox{
    \hbox{
      \includegraphics[width=0.5\hsize, bb=0 0 826 438]{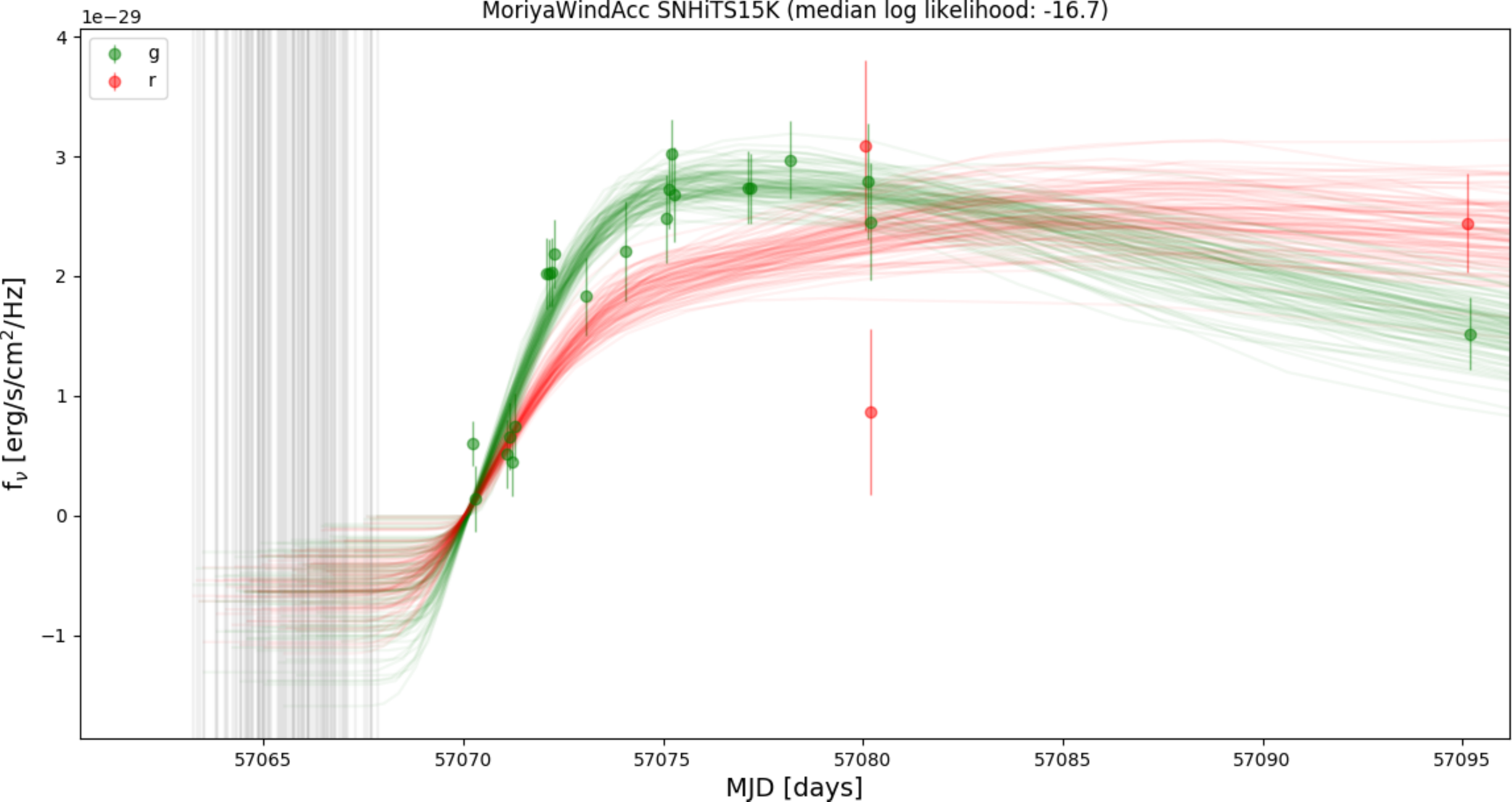}
      \includegraphics[width=0.5\hsize, bb=0 0 826 438]{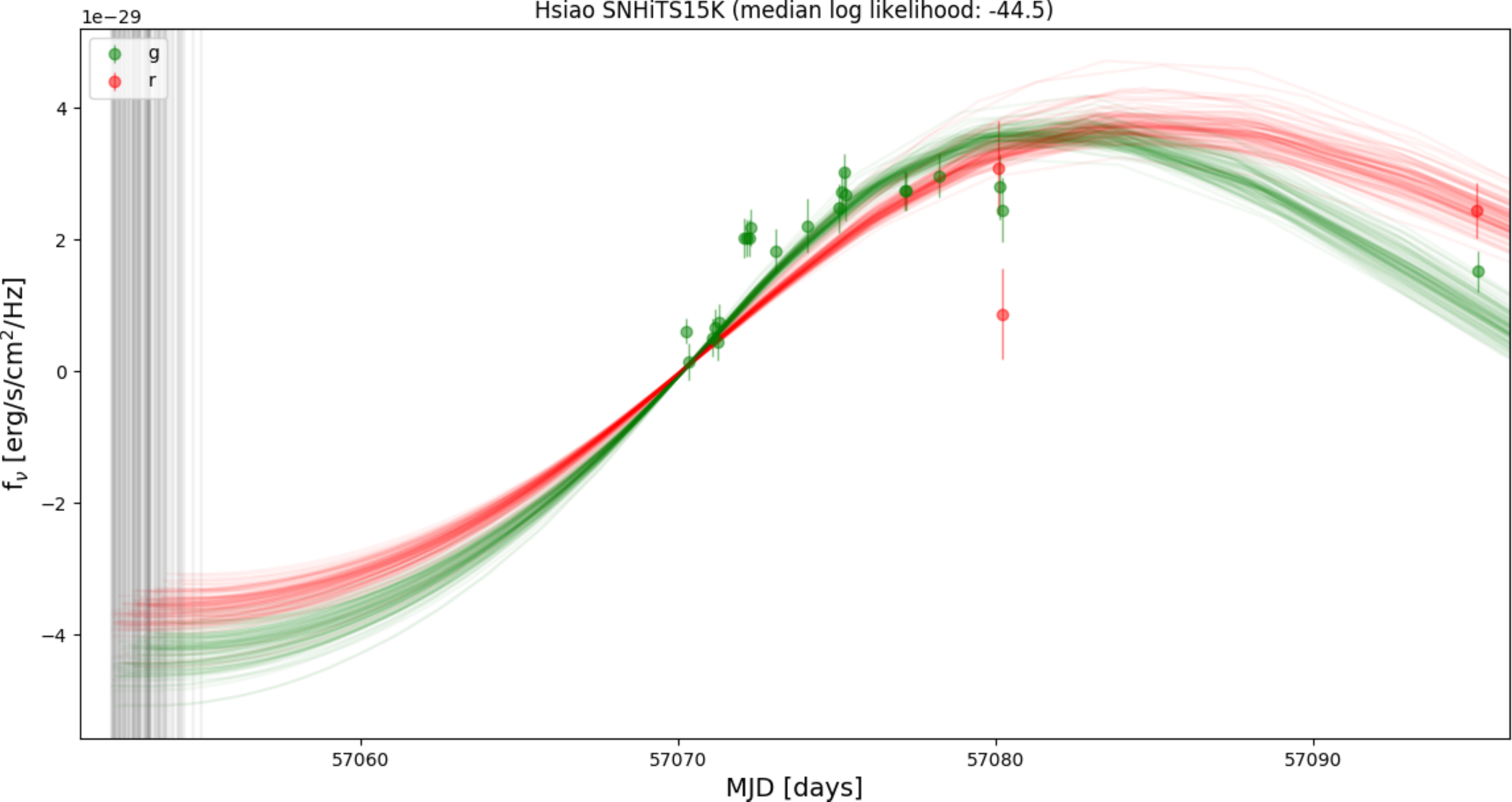}
    }
    \hbox{
      \includegraphics[width=0.5\hsize, bb=0 0 826 438]{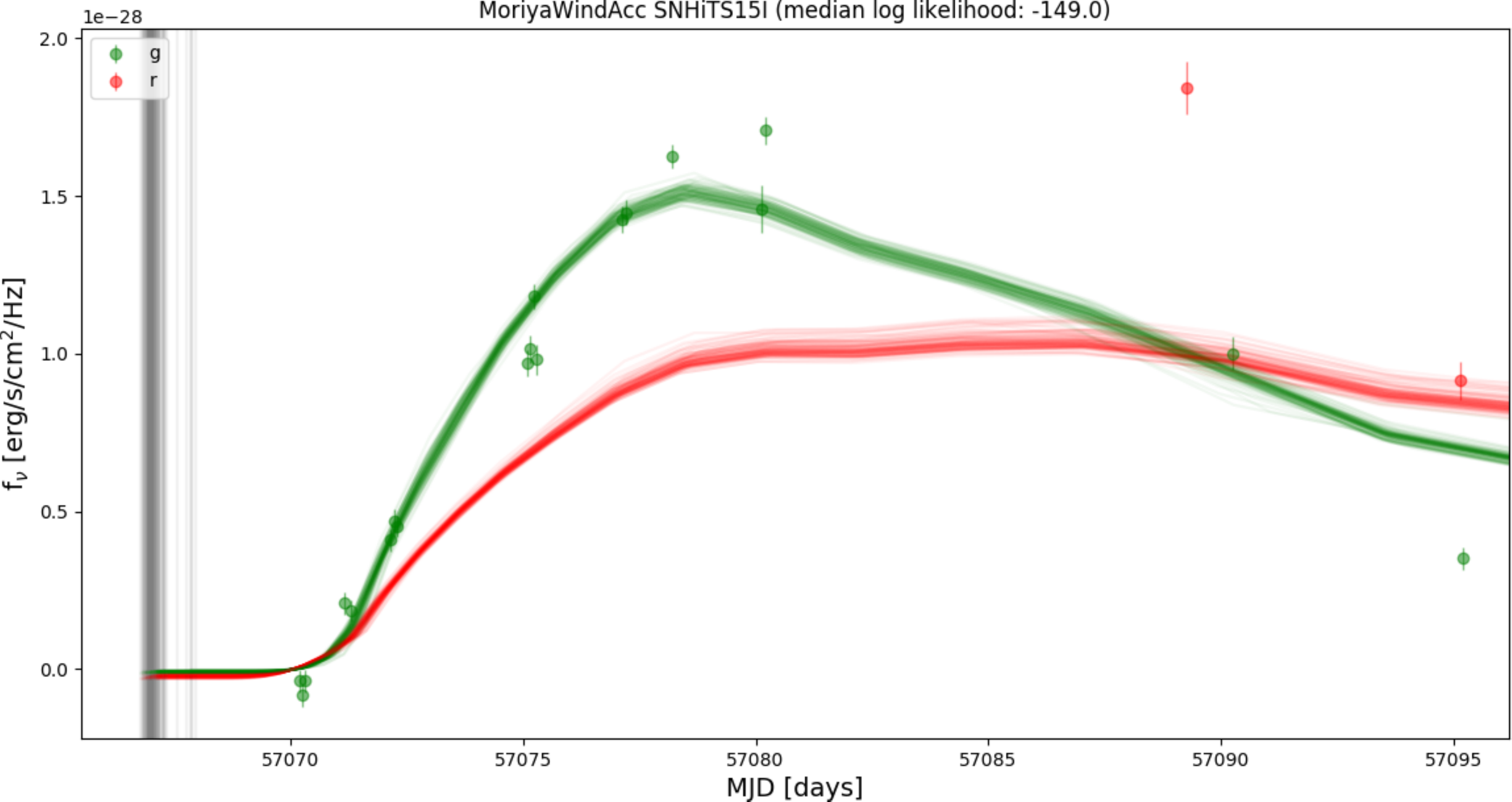}
      \includegraphics[width=0.5\hsize, bb=0 0 826 438]{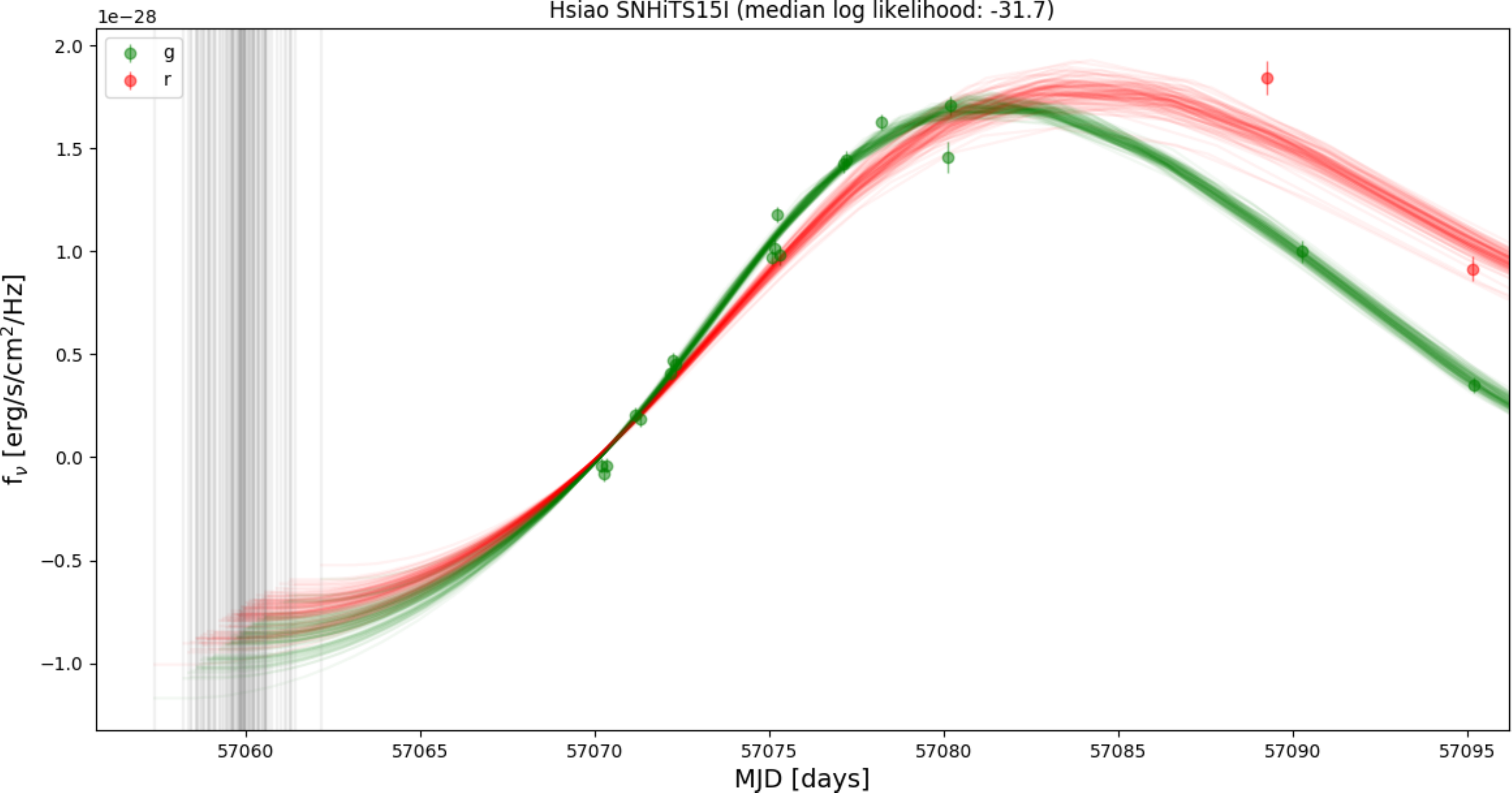}
    }
  }
  \caption{Observations of SNHiTS15K (top row) and SNHiTS15I (bottom
    row) with models sampled from the posterior distribution of
    parameters for the RSG SN II models (left column) and SN I models
    (right column). SNHiTS15K is best fitted by a SN II model, with a
    median log--likelihood of -16.7 for the SN II model vs -44.5 for
    the SN I model; while SNHiTS15I is best fitted by a SN I model, with
    a median log--likelihood of -149 for the SN II model and -31.7 for
    the SN I model.}
  \label{supfig:classification_examples}
\end{figure*}

\begin{figure}[ht!]
\hbox{
\includegraphics[width=0.5\hsize, bb=0 0 450 350]{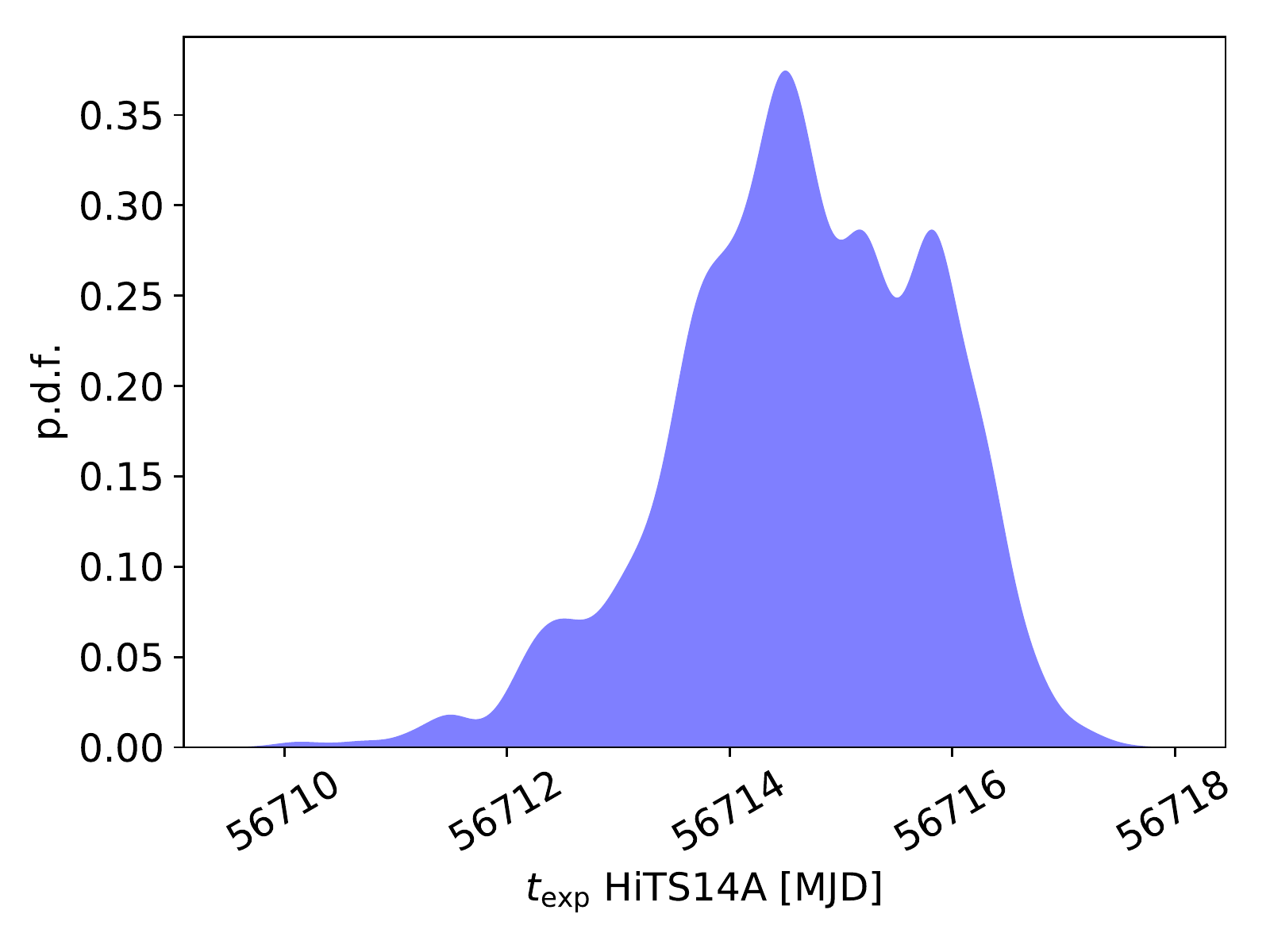}
\includegraphics[width=0.5\hsize, bb=0 0 450 350]{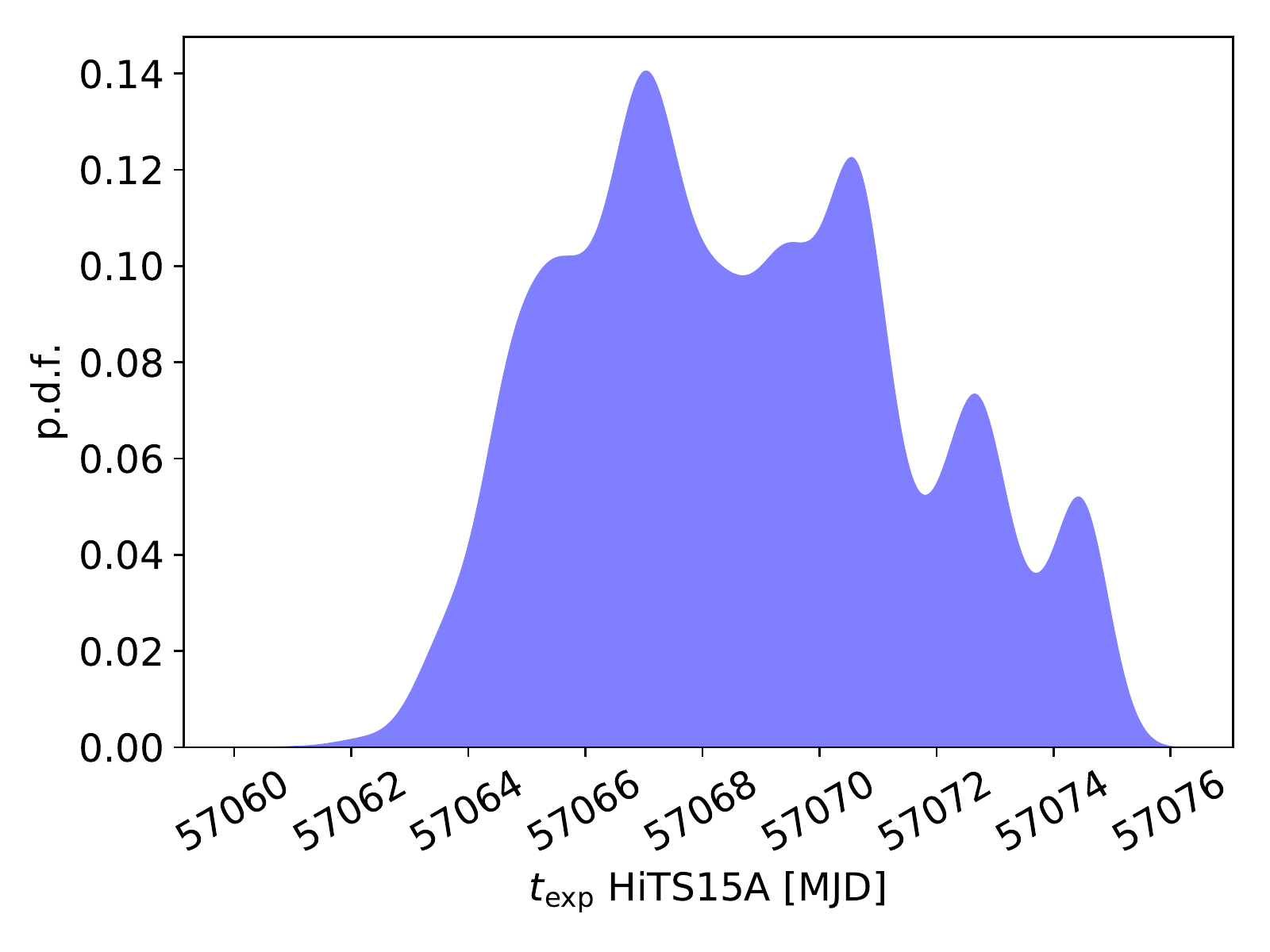}
}
\caption{Same as Figure~\ref{supfig:mdotbeta}, but for the distribution of median explosion times for the HiTS14A (l.h.s.) and HiTS15A (r.h.s.) campaigns.}
\label{supfig:texp}
\end{figure}

\begin{figure}[ht!]
  \hbox{
    \includegraphics[width=0.5\hsize, bb=0 0 450 350]{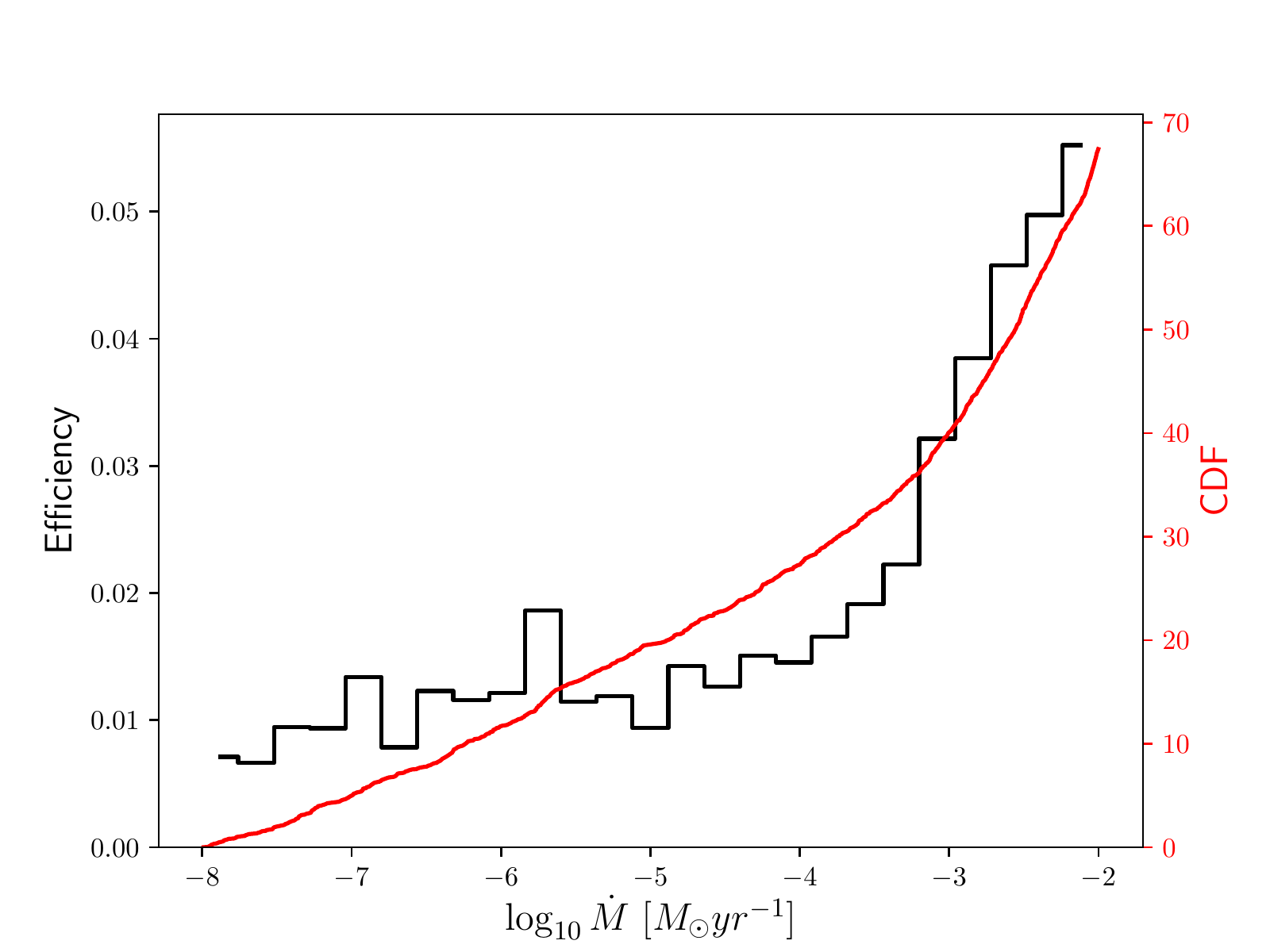}
    \includegraphics[width=0.5\hsize, bb=0 0 450 350]{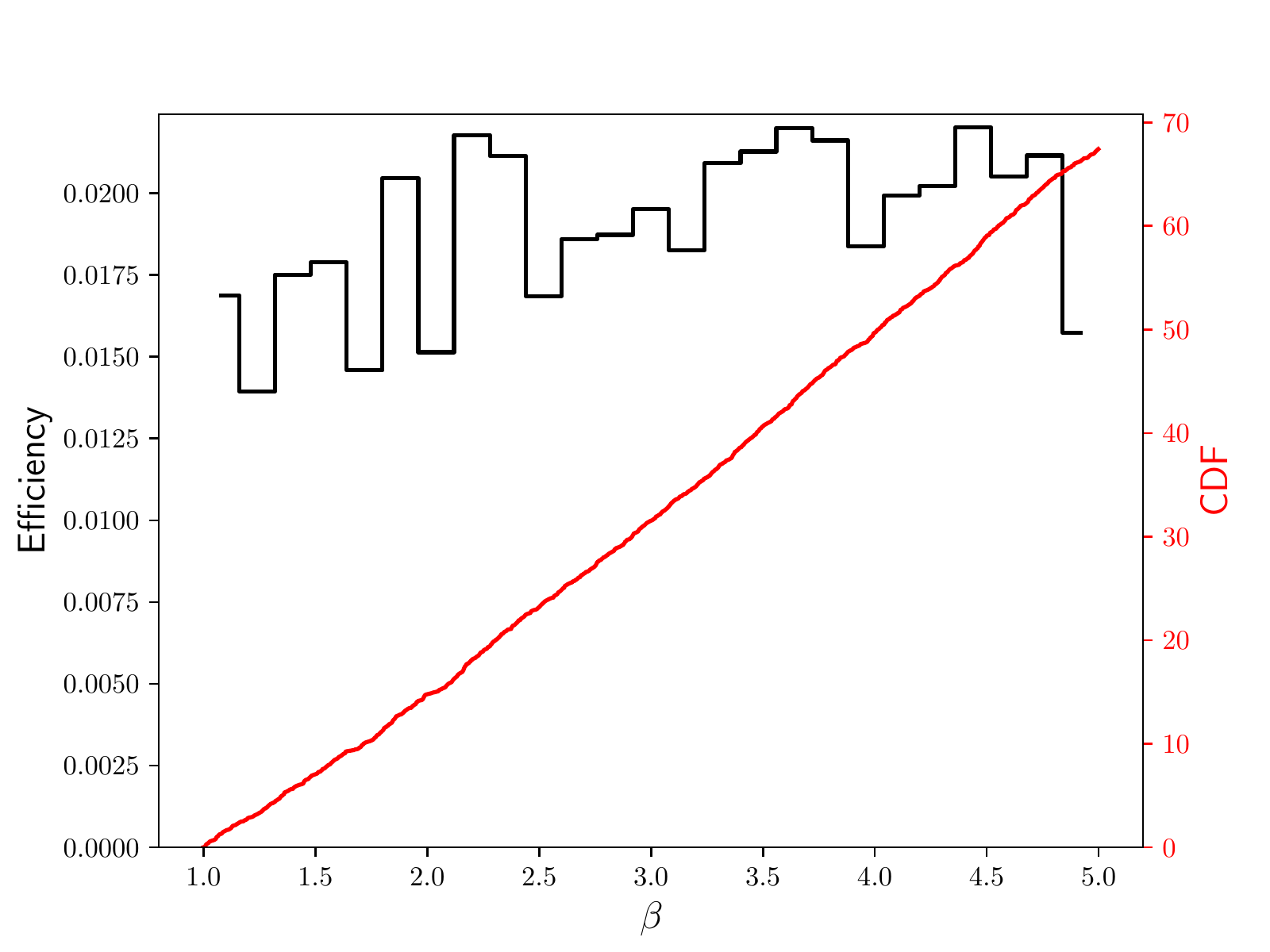}
  }
  \caption{Detection efficiency as a function of $\dot M$ and $\beta$
    predicted by our simulations using a realistic distribution of
    redshifts and assuming the same depth and cadence as in our
    observations.}
  \label{supfig:eff}
\end{figure}

\begin{figure}[ht!]
\hbox{
\includegraphics[width=0.5\hsize, bb=0 0 450 350]{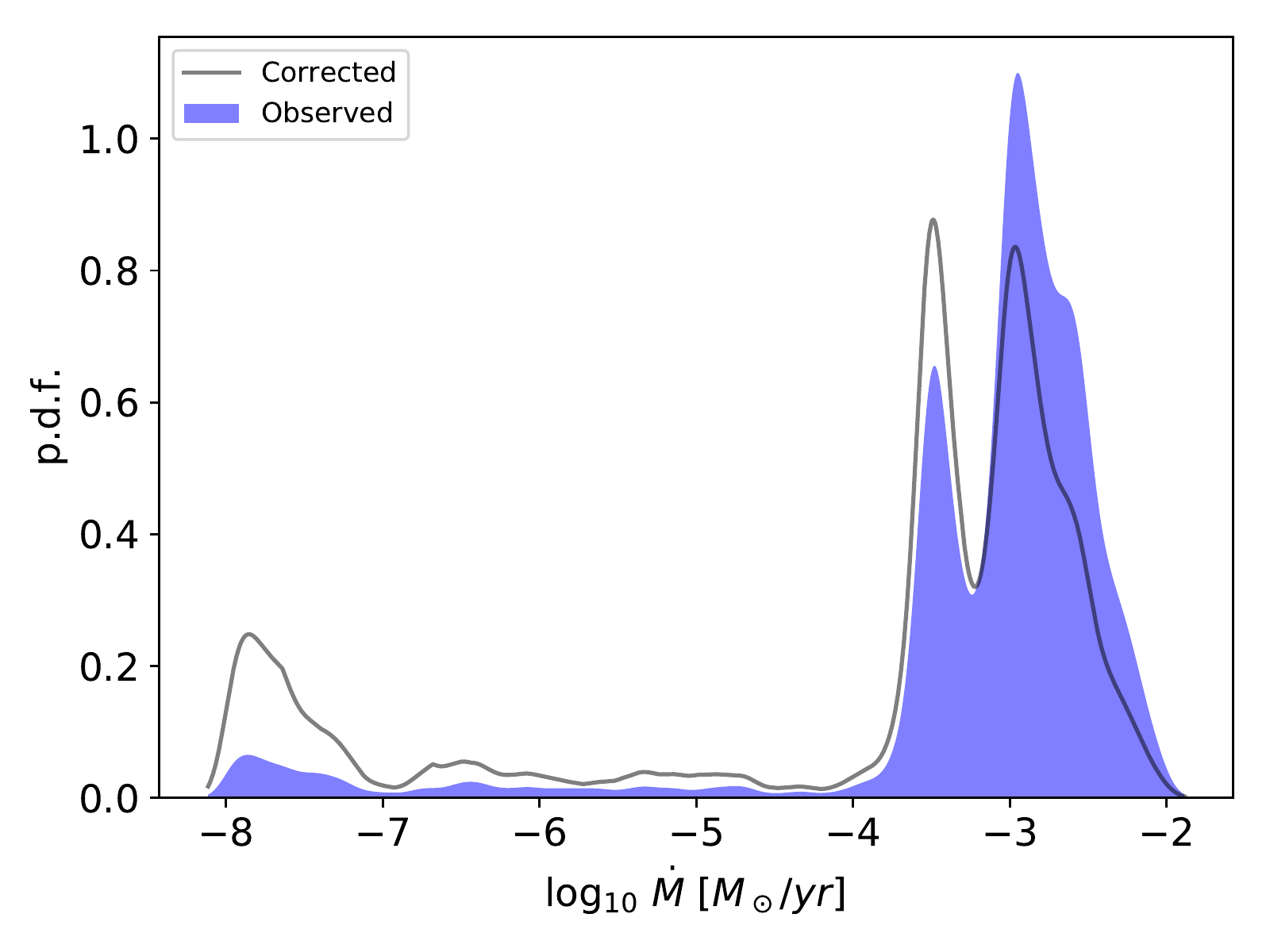}
\includegraphics[width=0.5\hsize, bb=0 0 450 350]{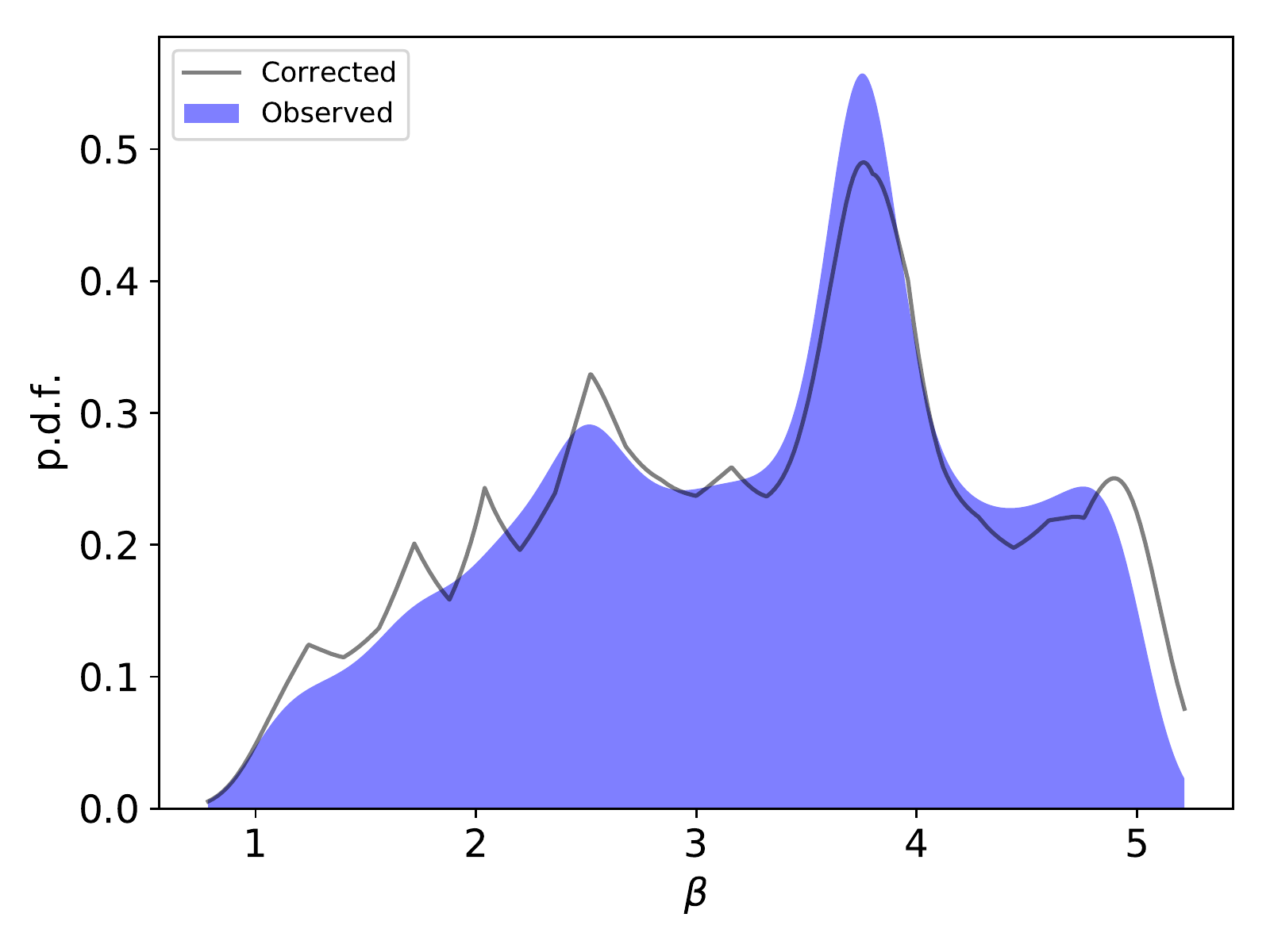}
}
\caption{Kernel density estimation of the sum of the sample posteriors
  marginalized over the mass loss rate $\dot M$ (l.h.s.) and wind
  acceleration parameter $\beta$ (r.h.s.) for the 26 SNe in the
  sample, using Silverman's rule to estimate the kernel width. The
  density estimations are corrected by the detection efficiencies
  shown in Figure~\ref{supfig:eff}.}
\label{supfig:mdotbeta}
\end{figure}

\begin{figure}[ht!]
\hbox{
\includegraphics[width=0.5\hsize, bb=0 0 450 350]{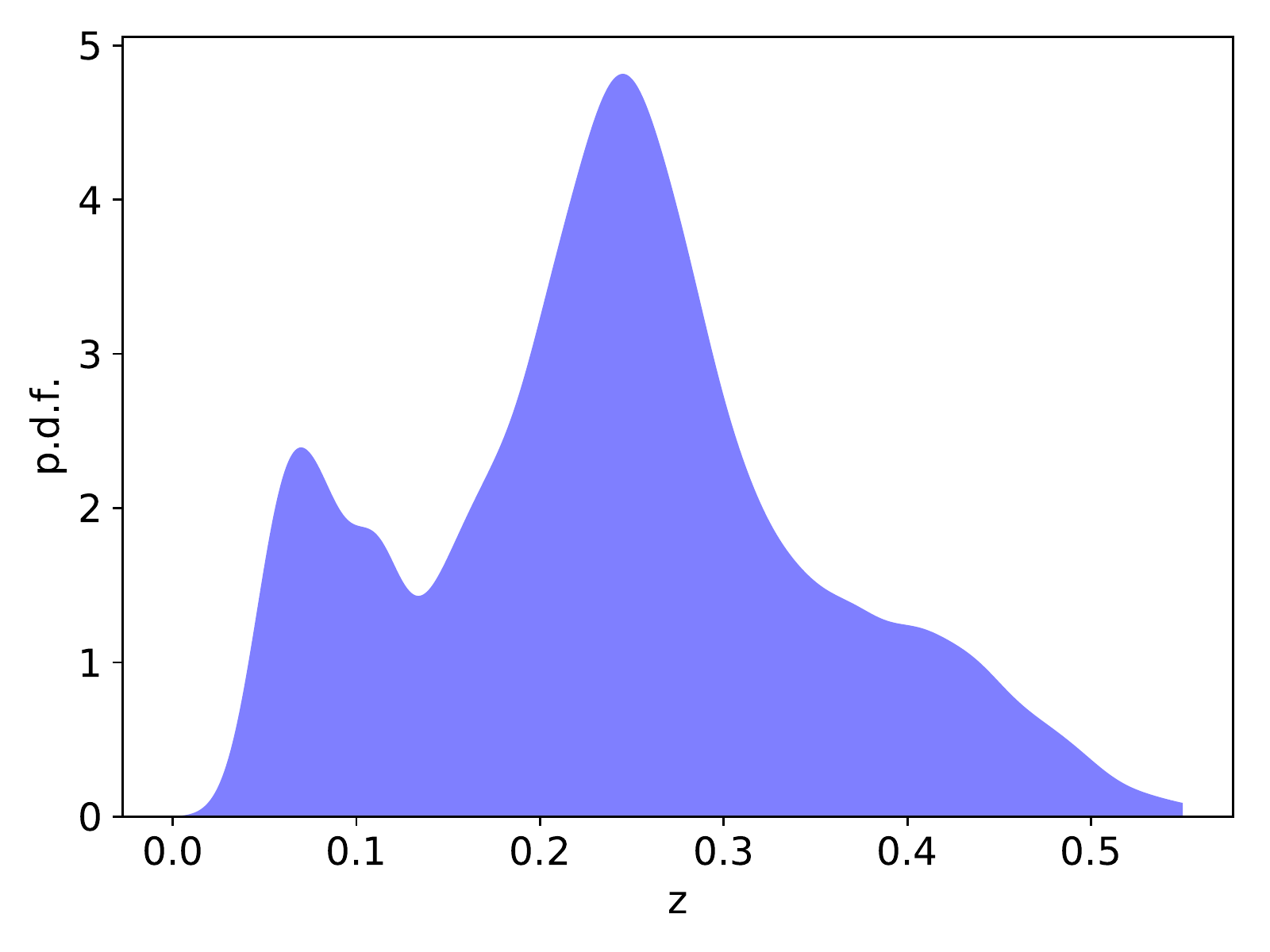}
\includegraphics[width=0.5\hsize, bb=0 0 450 350]{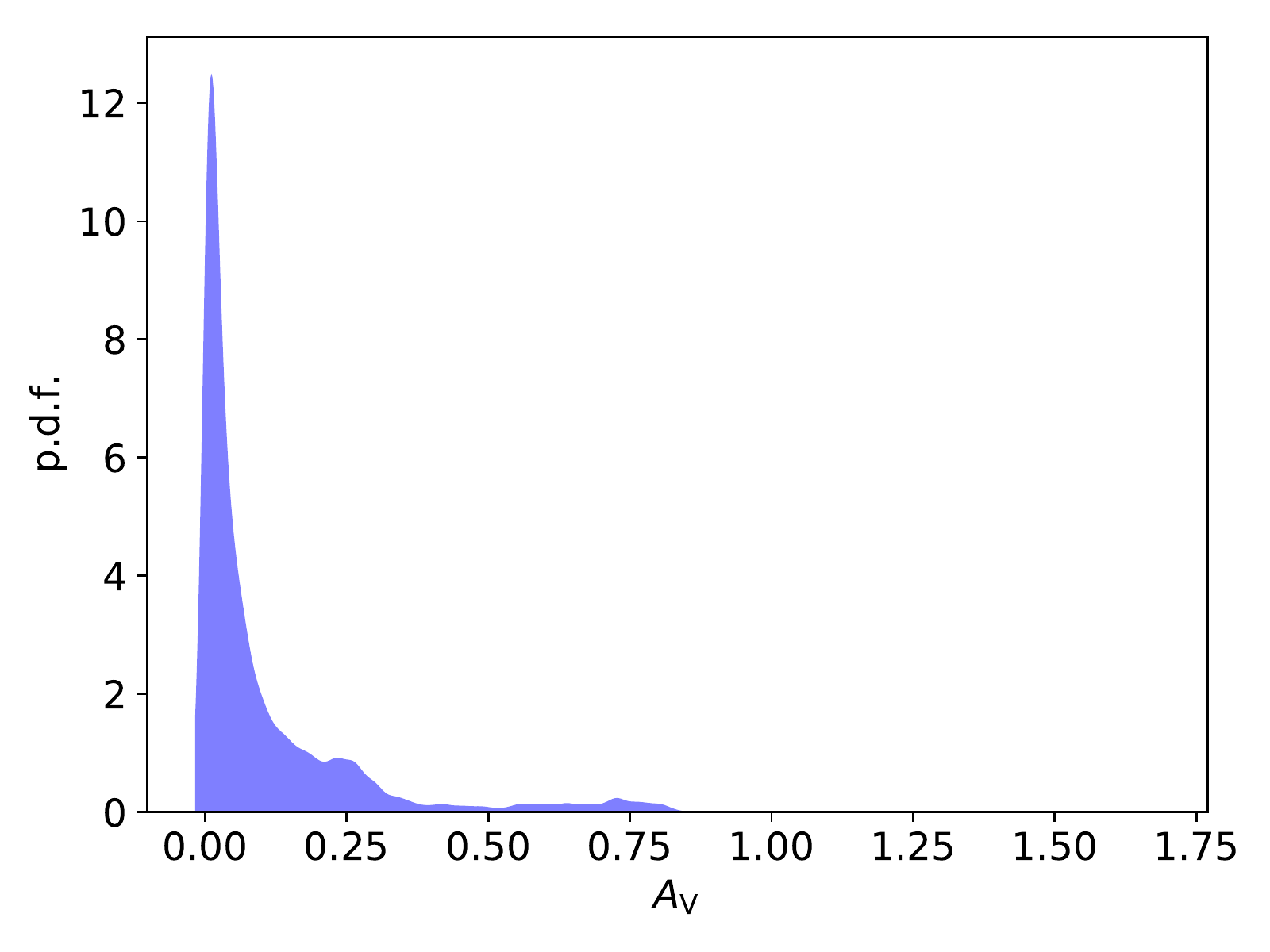}
}
\caption{Same as Figure~\ref{supfig:mdotbeta}, but for the distribution
  of redshift (l.h.s.) and attenuation (r.h.s.). Note that the
  apparent redshift bimodality is an artifact of the kernel density
  estimation.  Using a Hartigans' Dip test of Unimodality \cite{HH85}
  with the inferred median redshift values we could not reject
  unimodality with a $p$--value of 0.84}
\label{supfig:zAv}
\end{figure}

\begin{figure}[ht!]
  \begin{centering}
    \includegraphics[width=0.7\hsize, bb=0 0 450 350]{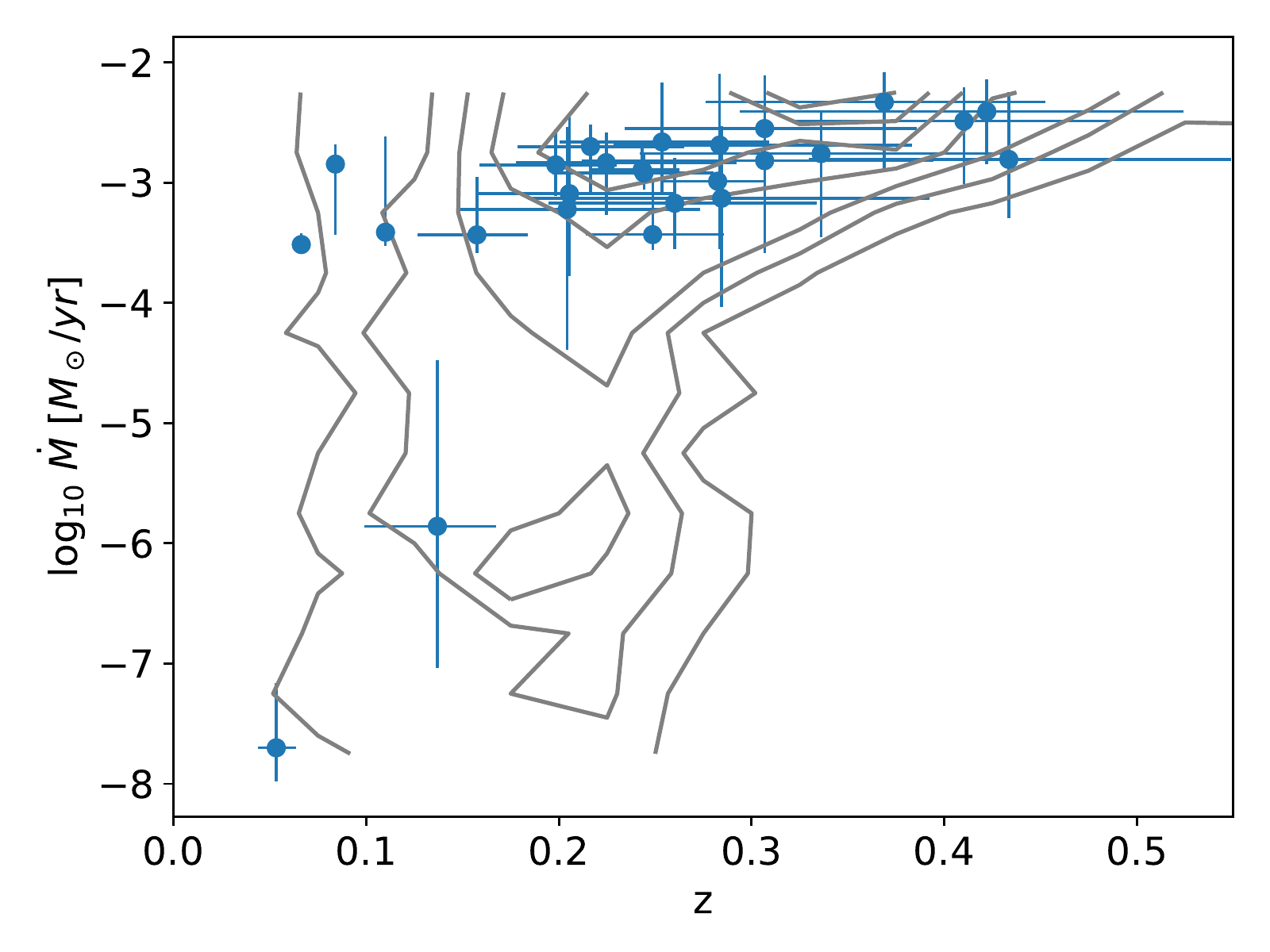}
    \caption{Relation between inferred mass loss rate $\dot M$ and
      redshift for the sample of 26 early SNe II candidates from
      HiTS. The error bars correspond to the percentiles 5, 50 and 95
      of the marginalized posterior distributions. We show for
      comparison the predicted distribution of $\log_{10} \dot M$ and
      redshift, as linearly spaced number density contours, in the
      case of a uniform distribution of $\log_{10} \dot M$ values (see
      text for more details).}
    \label{supfig:z} 
  \end{centering}
\end{figure}


\section*{Bibliography}

\begin{thebibliography}{10}
  
\bibitem{2009PASP..121.1395L} Law, Nicholas M., et al., The Palomar
  Transient Factory: System Overview, Performance, and First Results.
  \emph{Publ. Astron. Soc. Pac.}, {\bf 121}, 1395-1408 (2009).
  

\bibitem{2007PASA...24....1K} Keller, S.~C., et al., The SkyMapper
  Telescope and The Southern Sky Survey.  \emph{Publ. Astron. Soc. Aus.}, {\bf 24}, 1-12 (2007).


\bibitem{2010SPIE.7733E..0EK} Kaiser, Nick, et al., The Pan-STARRS
  wide-field optical/NIR imaging survey.  \emph{Soc. Phot. Inst. Eng.}, {\bf 7733}, 77330E
  (2010).


\bibitem{2011SPIE.8151E..1BK} Kim, Seung-Lee, et al., Wide-field
  telescope design for the KMTNet project.  \emph{Soc. Phot. Inst. Eng.}, {\bf 8151}, 81511B
  (2011).

\bibitem{2018PASP..130f4505T} Tonry, J.~L., et al., ATLAS: A
  High-cadence All-sky Survey System.  \emph{Publ. Astron. Soc. Pac.}, {\bf 130}, 064505 (2018).
  
\bibitem{2015AJ....150..150F} Flaugher, B., et al., The Dark Energy
  Camera.  \emph{Astron. J.}, {\bf 150}, 150 (2015).

\bibitem{2010AIPC.1279..120T} Takada, Masahiro, Subaru Hyper
  Suprime-Cam Project.  \emph{Proc. Am. Inst. Phys. Conf.}, {\bf 1279}, 120 (2010).
  
\bibitem{2017NatAs...1E..71B} Bellm, Eric \& Kulkarni, Shrinivas The
  unblinking eye on the sky.  \emph{Nat. Astron.}, {\bf 1}, 0071 (2017).
  
\bibitem{2009arXiv0912.0201L} {LSST Science Collaboration}, {\it
  et~al.\/}, Preprint at \url{http://arxiv.org/abs/0912.0201} (2009).

\bibitem{2016ApJ...832..155F} F{\"o}rster, F., et al., The High
  Cadence Transient Survey (HITS). I. Survey Design and Supernova
  Shock Breakout Constraints.  \emph{Astrophys. J.}, {\bf 832}, 155 (2016).

\bibitem{2009MNRAS.395.1409S} Smartt, S.~J., Eldridge, J.~J., Crockett,
  R.~M.,
  \& Maund, J.~R.,  The death of massive stars - I. Observational constraints on the progenitors of Type II-P supernovae.  \emph{Mon. Not. R. Astron. Soc.},  {\bf 395}, 1409 (2009).

\bibitem{2018MNRAS.474.2116D} Davies, Ben \& Beasor, Emma R., The
  initial masses of the red supergiant progenitors to Type II
  supernovae.  \emph{Mon. Not. R. Astron. Soc.}, {\bf 474}, 2116 (2018).
  
\bibitem{2016ARNPS..66..341J} Janka, Hans-Thomas, Melson, Tobias, \&
  Summa, Alexander, Physics of Core-Collapse Supernovae in Three
  Dimensions: A Sneak Preview.  \emph{Annu. Rev. Nucl. Phys.}, {\bf 66}, 341 (2016).

\bibitem{1978ApJ...225L.133F} Falk, S.~W., Shock steepening and prompt
  thermal emission in supernovae.  \emph{Astrophys. J.}, {\bf 225}, L133 (1978).
  
\bibitem{1992ApJ...393..742E} Ensman, Lisa \& Burrows, Adam, Shock
  breakout in SN 1987A.  \emph{Astrophys. J.}, {\bf 393}, 742 (1992).

\bibitem{2017hsn..book..967W} Waxman, Eli \& Katz, Boaz, Shock
  Breakout Theory, \emph{Handbook of Supernovae}, 967 (2017).
    
\bibitem{2011ApJS..193...20T} Tominaga, N., et al., Shock Breakout in
  Type II Plateau Supernovae: Prospects for High-Redshift Supernova
  Surveys.  \emph{Astrophys. J. Suppl. Ser.}, {\bf 193}, 20 (2011).
  
\bibitem{2010ApJ...724.1396O} Ofek, E.~O., et al., Supernova PTF 09UJ:
  A Possible Shock Breakout from a Dense Circumstellar Wind.  \emph{Astrophys. J.},
  {\bf 724}, 1396 (2010).

\bibitem{2015ApJ...804...28G} Gezari, S., et al., GALEX Detection of
  Shock Breakout in Type IIP Supernova PS1-13arp: Implications for the
  Progenitor Star Wind.  \emph{Astrophys. J.}, {\bf 804}, 28 (2015).
  
\bibitem{2017MNRAS.469L.108M} Moriya, Takashi J., Yoon, Sung-Chul,
  Gr{\"a}fener, G{\"o}tz, \& Blinnikov, Sergei I., Immediate dense
  circumstellar environment of supernova progenitors caused by wind
  acceleration: its effect on supernova light curves.  \emph{Mon. Not. R. Astron. Soc.}, {\bf
    469}, L108 (2017).
    
\bibitem{2018MNRAS.476.2840M} Moriya, Takashi J., F{\"o}rster,
  Francisco, Yoon, Sung-Chul, Gr{\"a}fener, G{\"o}tz, \& Blinnikov,
  Sergei I., Type IIP supernova light curves affected by the
  acceleration of red supergiant winds.  \emph{Mon. Not. R. Astron. Soc.}, {\bf 476}, 2840
  (2018).

\bibitem{2017NatPh..13..510Y} Yaron, O., et al., Confined dense
  circumstellar material surrounding a regular type II supernova.
  \emph{Nat. Phys.}, {\bf 13}, 510 (2017).

\bibitem{2017A&A...605A..83D} Dessart, Luc, John Hillier, D., \&
  Audit, Edouard, Explosion of red-supergiant stars: Influence of the
  atmospheric structure on shock breakout and early-time supernova
  radiation.  \emph{Astron. Astrophys.}, {\bf 605}, A83 (2017).

\bibitem{2015MNRAS.451.2212G} Gonz{\'a}lez-Gait{\'a}n, S., et al., The
  rise-time of Type II supernovae.  \emph{Mon. Not. R. Astron. Soc.}, {\bf 451}, 2212 (2015).

\bibitem{2016ApJ...828..111R} Rubin, Adam \& Gal-Yam, Avishay,
  Unsupervised Clustering of Type II Supernova Light Curves.  \emph{Astrophys. J.},
  {\bf 828}, 111 (2016).
  
\bibitem{2017ApJ...848....8R} Rubin, Adam \& Gal-Yam, Avishay,
  Exploring the Efficacy and Limitations of Shock-cooling Models: New
  Analysis of Type II Supernovae Observed by the Kepler Mission.  \emph{Astrophys. J.},
  {\bf 848}, 8 (2017).
  
\bibitem{2017ApJ...838...28M} Morozova, Viktoriya, Piro, Anthony L.,
  \& Valenti, Stefano, Unifying Type II Supernova Light Curves with
  Dense Circumstellar Material.  \emph{Astrophys. J.}, {\bf 838}, 28 (2017).
  
\bibitem{2008Sci...321..223S} Schawinski, Kevin, et al., Supernova
  Shock Breakout from a Red Supergiant.  \emph{Sci}, {\bf 321}, 223 (2008).

\bibitem{2008ApJ...683L.131G} Gezari, Suvi, et al.,  Probing Shock Breakout
  with Serendipitous GALEX Detections of Two SNLS Type II-P Supernovae.  \emph{Astrophys. J.},
  {\bf 683}, L131 (2008). 
  
\bibitem{2016ApJ...820...23G} Garnavich, P.~M., et al., Shock Breakout
  and Early Light Curves of Type II-P Supernovae Observed with Kepler.
  \emph{Astrophys. J.}, {\bf 820}, 23 (2016).

\bibitem{2016ApJ...820...57G} Ganot, Noam, et al.,  The Detection Rate of
  Early UV Emission from Supernovae: A Dedicated Galex/PTF Survey and
  Calibrated Theoretical Estimates.  \emph{Astrophys. J.},  {\bf 820}, 57 (2016).

\bibitem{2016ApJ...819....5T} Tanaka, Masaomi, et al.,  Rapidly Rising
  Transients from the Subaru Hyper Suprime-Cam Transient Survey.  \emph{Astrophys. J.},  {\bf
    819}, 5 (2016). 

\bibitem{2018Natur.554..497B} Bersten, M.~C., et al., A surge of light
  at the birth of a supernova.  \emph{Nat.}, {\bf 554}, 497 (2018).
  
\bibitem{2015A&A...579A..40S} Smartt, S.~J., et al., PESSTO: survey
  description and products from the first data release by the Public
  ESO Spectroscopic Survey of Transient Objects.  \emph{Astron. Astrophys.}, {\bf 579}, A40
  (2015).
  
\bibitem{2016MNRAS.457..328L} Lyman, J.~D., et al., Bolometric light
  curves and explosion parameters of 38 stripped-envelope
  core-collapse supernovae.  \emph{Mon. Not. R. Astron. Soc.}, {\bf 457}, 328 (2016).

\bibitem{2007ApJ...663.1187H} Hsiao, E.~Y., et al., K-Corrections and
  Spectral Templates of Type Ia Supernovae.  \emph{Astrophys. J.}, {\bf 663}, 1187
  (2007).

\bibitem{2007ApJ...666.1093Q} Quimby, Robert M., et al.,  SN 2006bp:
  Probing the Shock Breakout of a Type II-P Supernova.  \emph{Astrophys. J.},  {\bf 666}, 1093
  (2007). 

\bibitem{2011ApJS..192....3P} Paxton, Bill, et al., Modules for
  Experiments in Stellar Astrophysics (MESA).  \emph{Astrophys. J. Suppl. Ser.}, {\bf 192}, 3
  (2011).

\bibitem{2013ApJS..208....4P} Paxton, Bill, et al., Modules for
  Experiments in Stellar Astrophysics (MESA): Planets, Oscillations,
  Rotation, and Massive Stars.  \emph{Astrophys. J. Suppl. Ser.}, {\bf 208}, 4 (2013).

\bibitem{2015ApJS..220...15P} Paxton, Bill, et al.,  Modules for
  Experiments in Stellar Astrophysics (MESA): Binaries, Pulsations, and
  Explosions.  \emph{Astrophys. J. Suppl. Ser.},  {\bf 220}, 15 (2015). 

\bibitem{1996A&A...305..171P} Puls, J., et al., O-star mass-loss and
  wind momentum rates in the Galaxy and the Magellanic Clouds
  Observations and theoretical predictions.  \emph{Astron. Astrophys.}, {\bf 305}, 171
  (1996).

  
\bibitem{2010ASPC..425..181B} Bennett, P.~D.,  Chromospheres and Winds of
  Red Supergiants: An Empirical Look at Outer Atmospheric Structure.  \emph{ASPC},
  {\bf 425}, 181 (2010). 

\bibitem{2004MNRAS.355.1348M} Marshall, Jonathan R., et al.,  Asymptotic
  giant branch superwind speed at low metallicity.  \emph{Mon. Not. R. Astron. Soc.},  {\bf 355}, 1348
  (2004).

\bibitem{1996ApJ...466..979B} Baade, Robert, et al.,  The Wind Outflow of
  zeta Aurigae: A Model Revision Using Hubble Space Telescope Spectra.  \emph{Astrophys. J.},
  {\bf 466}, 979 (1996). 
  
\bibitem{2011A&A...526A.156M} Mauron, N.~\& Josselin, E., The
  mass-loss rates of red supergiants and the de Jager prescription.
  \emph{Astron. Astrophys.}, {\bf 526}, A156 (2011).

\bibitem{2017MNRAS.465..403G} Goldman, Steven R., et al.,  The wind speeds,
  dust content, and mass-loss rates of evolved AGB and RSG stars at varying
  metallicity.  \emph{Mon. Not. R. Astron. Soc.},  {\bf 465}, 403 (2017). 

\bibitem{2018arXiv180407312M} Morozova, Viktoriya \& Stone, James M.,
  Theoretical X-ray light curves of young SNe II: the example of SN
  2013ej. Preprint at \url{http://arxiv.org/abs/1804.07312} (2018).

\bibitem{1998ApJ...496..454B} Blinnikov, S.~I., Eastman, R., Bartunov,
  O.~S., Popolitov, V.~A., \& Woosley, S.~E., A Comparative Modeling
  of Supernova 1993J.  \emph{Astrophys. J.}, {\bf 496}, 454 (1998).
    
\bibitem{2000ApJ...532.1132B} Blinnikov, Sergei, Lundqvist, Peter,
  Bartunov, Oleg, Nomoto, Ken'ichi, \& Iwamoto, Koichi, Radiation
  Hydrodynamics of SN 1987A. I. Global Analysis of the Light Curve for
  the First 4 Months.  \emph{Astrophys. J.}, {\bf 532}, 1132 (2000).
  
\bibitem{2006A&A...453..229B} Blinnikov, S.~I., et al., Theoretical
  light curves for deflagration models of type Ia supernova.  \emph{Astron. Astrophys.},
  {\bf 453}, 229 (2006).

\bibitem{2014Natur.509..471G} Gal-Yam, Avishay, et al., A
  Wolf-Rayet-like progenitor of SN 2013cu from spectral observations
  of a stellar wind.  \emph{Nat.}, {\bf 509}, 471 (2014).
  
\bibitem{2014A&A...572L..11G} Groh, Jose H., Early-time spectra of
  supernovae and their precursor winds. The luminous blue
  variable/yellow hypergiant progenitor of SN 2013cu.  \emph{Astron. Astrophys.}, {\bf
    572}, L11 (2014).

\bibitem{2016MNRAS.455..112G} Gr{\"a}fener, G.~\& Vink, J.~S.,
  Light-travel-time diagnostics in early supernova spectra:
  substantial mass-loss of the IIb progenitor of SN 2013cu through a
  superwind.  \emph{Mon. Not. R. Astron. Soc.}, {\bf 455}, 112 (2016).
  
\bibitem{1985A&A...147..103S} Schroeder, K.-P., A study of ultraviolet
  spectra of Zeta Aurigae/VV Cephei systems. VII - Chromospheric
  density distribution and wind acceleration region.  \emph{Astron. Astrophys.}, {\bf 147},
  103 (1985).

\bibitem{2017Natur.548..310O} Ohnaka, K., Weigelt, G., \& Hofmann,
  K.-H., Vigorous atmospheric motion in the red supergiant star
  Antares.  \emph{Nat.}, {\bf 548}, 310 (2017).

\bibitem{2014AJ....148..107R} Rodr{\'{\i}}guez, {\'O}smar,
  Clocchiatti, Alejandro, \& Hamuy, Mario, Photospheric Magnitude
  Diagrams for Type II Supernovae: A Promising Tool to Compute
  Distances.  \emph{Astron. J.}, {\bf 148}, 107 (2014).
  

\end{thebibliography}

\begin{thebibliography}{10}

\bibitem{2014ATel.5957....1W} Walton, N., et al.,  PESSTO spectroscopic
  classification of optical transients.  \emph{Astron. Tel.},  {\bf 5957}, (2014). 

\bibitem{2014ATel.5970....1W} Walton, N., et al.,  PESSTO spectroscopic
  classification of optical transients.  \emph{Astron. Tel.},  {\bf 5970}, (2014). 

\bibitem{2015ATel.7144....1L} Le Guillou, L., et al.,  PESSTO spectroscopic
  classification of optical transients.  \emph{Astron. Tel.},  {\bf 7144}, (2015). 

\bibitem{2015ATel.7154....1B} Baumont, S., et al.,  PESSTO spectroscopic
  classification of optical transients.  \emph{Astron. Tel.},  {\bf 7154}, (2015). 
  
\bibitem{2015ATel.7291....1F} Forster, F., et al.,  Optical spectra of
  SNHiTS15al, SNHiTS15be, SNHiTS15bs and SNHiTS15by.  \emph{Astron. Tel.},  {\bf 7291},
  (2015). 

\bibitem{2015ATel.7246....1P} Pignata, G., et al.,  Optical spectroscopy of
  SNHiTS15aw.  \emph{Astron. Tel.},  {\bf 7246}, (2015).

\bibitem{2015ATel.7164....1A} Anderson, J., et al.,  Optical spectrosopy of
  SNHiTS15ad (Gabriela).  \emph{Astron. Tel.},  {\bf 7164}, (2015). 

\bibitem{2015ATel.7335....1A} Anderson, J., et al.,  Optical spectrosopy of
  HiTS supernovae.  \emph{Astron. Tel.},  {\bf 7335}, (2015). 

\bibitem{2014ATel.6014....1A} Anderson, J., et al., FORS2
  spectroscopic classification of DECam SN candidates.  \emph{Astron. Tel.}, {\bf
    6014}, (2014).

\bibitem{2015ATel.7162....1A} Anderson, J., et al.,  Optical spectrosopy of
  SNHiTS15D (Daniela) and SNHiTS15P (Rosemary).  \emph{Astron. Tel.},  {\bf 7162}, (2015).

\bibitem{2007ApJ...666.1024B} Blondin, St{\'e}phane \& Tonry, John L.,
  Determining the Type, Redshift, and Age of a Supernova Spectrum.
  \emph{Astrophys. J.}, {\bf 666}, 1024 (2007).
  
\bibitem{2010PASP..122.1236M} Mighell, Kenneth John, CRBLASTER: A
  Parallel-Processing Computational Framework for Embarrassingly
  Parallel Image-Analysis Algorithms.  \emph{Publ. Astron. Soc. Pac.}, {\bf 122}, 1236 (2010).

\bibitem{2016arXiv161205243F} Flewelling, H.~A., et al., The
  Pan-STARRS1 Database and Data Products.  Preprint at
  \url{http://arxiv.org/abs/1612.05243} (2016).

\bibitem{2010CAMCS...5...65G} Goodman, Jonathan
  \& Weare, Jonathan,  Ensemble samplers with affine invariance.  \emph{Comm. Appl. Math. Comp. Sci.},  {\bf 5}, 65 (2010). 

\bibitem{2013PASP..125..306F} Foreman-Mackey, Daniel, Hogg, David W.,
  Lang, Dustin, \& Goodman, Jonathan, emcee: The MCMC Hammer.  \emph{Publ. Astron. Soc. Pac.},
  {\bf 125}, 306 (2013).
  
\bibitem{1986desd.book.....S} Silverman, B.~W., Density estimation for
  statistics and data analysis. \emph{Mon. Stat. Appl. Prob.} (1986).

\bibitem{HH85} J.~A. {Hartigan}, P.~M. {Hartigan}, The Dip Test of
  Unimodality. \emph{Ann. Statist.}, {\bf 13}, 1, 70--84 (1985)

\end{thebibliography}
\end{document}